 \documentclass[preprint,aps,pre,showpacs,superscriptaddress,nofootinbib,amssymb]{revtex4}
 \usepackage{graphicx}
 \usepackage{dcolumn}
 \usepackage{bm}
 \def\non{\nonumber}
 \def\be{\begin{equation}}
 \def\ee{\end{equation}}
 \def\bea{\begin{eqnarray}}
 \def\eea{\end{eqnarray}}
 \def\bi{\begin{itemize}}
 \def\ei{\end{itemize}}

 \def\pathmix{.}

\begin{document}
 \preprint{\today}
 
 
 \title{ Gravitational Fugacity as the seed of self-organized violent relaxaton process toward Local Virial relation}

 \author{Yasuhide Sota \footnote{sota@cosmos.phys.ocha.ac.jp}}
   \affiliation{Department of Physics, Ochanomizu University,
 	       2-1-1 Ohtuka, Bunkyo, Tokyo 112-8610, Japan}
   \affiliation{Advanced Research Institute for Science and Engineering, 
                Waseda University, Ohkubo, Shinjuku--ku, Tokyo 169-8555, Japan}
 
 \author{Osamu Iguchi \footnote{osamu@phys.ocha.ac.jp}}
   \affiliation{Department of Physics, Ochanomizu University,
 	       2-1-1 Ohtuka, Bunkyo, Tokyo 112-8610, Japan}
 
 \author{Tohru Tashiro \footnote{tashiro@cosmos.phys.ocha.ac.jp}}
   \affiliation{Department of Physics, Ochanomizu University,
 	       2-1-1 Ohtuka, Bunkyo, Tokyo 112-8610, Japan}

 \author{Masahiro Morikawa \footnote{hiro@phys.ocha.ac.jp}}
   \affiliation{Department of Physics, Ochanomizu University,
                2-1-1 Ohtuka, Bunkyo, Tokyo 112-8610, Japan}

 \begin{abstract}

  We propose  the self-organized
 relaxation process  which drives a collisionless self-gravitating system (SGS)
 to the equilibrium
 state satisfying local virial (LV) relation.
 During the violent relaxation process, 
 particles can move widely within the time
 interval as short as a few free fall times, because
 of the effective potential oscillations. 
 Since such particle movement causes further
 potential oscillations,
 it is expected
 that the system approaches the critical state where
  such particle activities, which we call gravitational fugacity,
 is  independent of the local position as much as possible.
 Here we  demonstrate  that 
 gravitational
 fugacity  can be described as the functional of the LV ratio, which
 means that the LV ratio is a key ingredient
 estimating the particle activities against gravitational potential.
 We also demonstrate that LV relation is attained if the LV ratio exceeds the crticaial value $b=1$
 everywhere in the bound region during the violent relaxation process.
 The local region which does not meet this criterion can be
  trapped into
 the pre-saturated  state.
  However, small phase-space perturbation can
 bring the inactive part into the LV critical state.
 \end{abstract}
 
 \maketitle

 \section{Introduction}
 
 It is well known that collisionless relaxation process plays a key role
 in driving the gravitational objects to  the equilibrium state 
 observed as elliptical galaxies or dark halos in our universe.
  Such  relaxation process has 
 been well analyzed with N-body simulations from the context of 
 the formation of density profile following $r^{1/4}$ law of elliptical galaxies \cite{Albada82,Aguilar91} or  dark halo formations after infalling from the cosmological background \cite{Navarro96,Navarro97,Taylor01}. 
 
 As a relaxation process toward such an equilibrium state,
 phase mixing and  violent relaxation has been proposed as a collisionless relaxation process,
 which, if completed, leads a self-gravitating system (SGS) to 
 the entropy maximum state called Linden-Bell distribution  \cite{Lynden67}.
 This distribution, however, cannot directly
 be applied to three dimensional open models, since it  has  an infinite mass
 and energy against the state attained through N-body simulations. 
 Actually, in numerical simulations, the violent relaxation
 is not completed and the state reaches the equilibrium state
 which cannot be described by the Linden-Bell statistics, 
  mainly due to the existence of 
 the particles with positive energy which escape infinitely \cite{Horthy91,Horthy93}.

 Then how can we characterize such a quasi-equilibrium state in an open system
 where particles can evaporate infinitely? Recently, we have
 numerically shown that   the bound
  states  after a cold collapse or cluster-cluster collisions are   virialized not only
 globally but also locally  for a wide range of initial conditions \cite{Sota04, Sota05,Osamu06}.
 Such a state with  the local virial (LV) relation is not a general solution for
 the stationary state of Vlasov equation. For example, Plummer's model
 is a unique solution satisfying the LV relation among the class of 
 polytropes. In addition, it is special among the
 class, since it has the unique solution with infinite extension of
 particles and with finite total mass \cite{Binney87}.
 From such viewpoints, Plummer's model was originally analyzed 
 by  Eddington \cite{Eddington16} and was  generalized 
 to  several families of both analytical and numerical solutions with anisotropic velocity dispersion \cite{Osamu06,
 Evans05}.

 Provoked by the remarkable characters, Eddington tried to show that
 Plummer's model is the local minimum of H function, but this approach was not fully successful \cite{Eddington16}. 
 So what kind of principle can characterize the LV relation as an 
 attractor?
 In general,  not all of the stationary state can be explained by the
 maximum entropy principle. For example, the non-equilibrium state is
 attained in the system sustained in the energy flow. In such open systems
 energy injection is balanced with the local energy dissipation.

 During the collisionless stage, potential oscillations play a key role
 in violent relaxation. Since such oscillations  are induced from the
 particle movements in the bound region, the activity of the
 particle against gravitational potentials is a key ingredient for
 the relaxation process.
 Here in this paper we propose the gravitational fugacity 
 which quantifies such particle activities against gravitational potential.
 We will also show that the fugacity can be described as the functional of
  LV ratio, which means that 
 LV ratio is the indicator estimating the local activity
 of particles against gravitational potential. This means that 
 the collisionless relaxation process of SGS is induced
 not from the entropy maximum principle minimizing the local temperature
 fluctuation, but rather from that minimizing local 
 fluctuations of particle fugacity.

 In Sec.\ref{sec:fugacity-def}, we define the local fugacity of particles as
 the generalization of local evaporation rate.
 In addition, we will  show that the fugacity can be the functional
 of LV ratio, which directly connect the LV ratio to particle activities
 against gravitational potential.
 We will also generalize it to  the case with
 the anisotropic Gaussian velocity distribution and show  that
 the local fugacity is sensitive to the LV ratio but
 is not so relevant with the anisotropy.
 In Sec.\ref{sec:fugacity-demo},
  we will numerically investigate the collisionless relaxation process
 for several initially spherical models and see 
 that LV relation is attained for the local region which experiences
 the high fugacity state where the fugacity is beyond the critical value. In Sec.\ref{sec:fugacity-demo2}, we will see characters of the LV relation 
 through several tests to find the criterion
 to achieve the LV critical state.
 Finally, in Sec.\ref{sec:conclusions}, we will summarize our analysis
 and comment on the  the analogy of  the LV relation with the critical
 state of self-organized criticality (SOC) or the role of LV relation
 from the viewpoints of superstatistics.
 . 
 \section{ LV relation as the state with constant fugacity 
 under global virial condition}
 \label{sec:fugacity-def}

 In general, gravitational relaxation is mainly divided into
 two processes. First is the collisionless relaxation process which is induced
 not through the particle-particle interaction but through the oscillation of
 gravitational potential. 
 Once the system
 is globally virialized, the relaxation process is ceased
 until two-body interactions are effective and the system begins
 to evolve toward two-body relaxation. The particle evaporation rate  defined as the rate of particles
 whose speed exceeds the local evaporation rate
 is a proper indicator for such collisional relaxation process, since the evaporation
 of particles occur only through the two-body interaction during the stationary collisionless stage.
 
 During the
 stage of 
 violent relaxation, on the other hand, evaporation rate is not necessarily a key ingredient for the relaxation process, since the two-body interaction is negligible during the period. In fact, as long as we examined, particles with positive energy  emerge 
 only at the moment of the maximum collapse
  with very low initial virial ratio. 
  However, during the violent relaxation, particle can be activated
  through the potential oscillations. The number of  particles which gain
  energy enough to move far away from the potential minimum increase
  even if they cannot escape infinitely from the gravitational center.
  
  In this section first we define the gravitational fugacity as the quantity which
  represents such particle activities against gravitational potential.
  Then we will show that the fugacity can be described as the 
  functional of LV ratio under the assumption that the local 
  velocity dispersion is isotropic.
  Finally we will also investigate how strongly the velocity anisotropy affects the
 relation between the gravitational fugacity and LV ratio. We will show that
 the anisotropy affects the gravitational fugacity  less
 sensitively than the LV ratio.  This means that the LV ratio
 becomes an effective indicator to estimate the particle activities
 against gravitational potential.
 
 \subsection{ The relation between  the LV ratio $b$  and the gravitational fugacity for isotropic Gaussian velocity distribution }
 \label{subsec:Gaussian}
  For simple demonstration, we first assume 
 that the velocity distribution
 is the locally Gaussian with the velocity dispersion $\sigma^2$
 which depends on position $\vec{r}$. 
 The Gaussianity of velocity distribution has been numerically supported
 at least in the central part of the bound region 
 for several numerical analyzes \cite{Merrall03,Osamu04}.
 In this case, the speed $v$ of each particle at the position $\vec{r}$ is governed by the following phase-space  density.
 \begin{equation}
 f(v,\vec{r}) = \rho(\vec{r})
 \left(\frac{3}{2\pi \sigma^2(\vec{r})}\right)^{3/2} 4\pi v^2
 \exp\left(-\frac{3v^2}{2\sigma^2(\vec{r})}\right),
 \label{gauss}
 \end{equation}
 where $\rho(\vec{r})$ is a mass density at $\vec{r}$.
 With the velocity dispersion $\sigma^2(\vec{r})$ and the local potential energy $\Phi(\vec{r})$, 
 the LV ratio is defined as follows \cite{Osamu06}:
 \begin{equation}
 b(\vec{r}) := -2\sigma^2(\vec{r})/\Phi(\vec{r}) .
 \label{eq1}
 \end{equation}
 
 Since the total energy of a particle evaporating from this system must be positive, we can assess the lowest speed $v_{\mbox{\scriptsize cr}}$ of the evaporating particles.
 \begin{equation}
 v_{\mbox{\scriptsize cr}} = \sqrt{-2\Phi(\vec{r})}
 \end{equation}
 
 In the local volume $dV$ at $\vec{r}$,
  the total mass of the particles whose speeds exceed $a v_{\mbox{\scriptsize cr}}$ with the constant value $a$ can be described as :
 \begin{equation}
 M_{a} \left(\vec{r}\right) = dV \rho \left(\vec{r}\right)R_{a}\left(\vec{r}\right),
 \end{equation}
 where $R_{a}$ is the rate of the particles  whose speeds exceed $a v_{\mbox{\scriptsize cr}}$, which is described as 
 \begin{equation}
 R_{a} \left(\vec{r}\right) = {1 \over \rho} \int_{a v_{\mbox{\tiny cr}}}^{\infty}f(v,\vec{r})dv \nonumber = 1-\mbox{Erf}\left[a \sqrt{-\frac{3\Phi}{\sigma^2}}\right]
 + 2 a  \sqrt{-\frac{3\Phi}{\pi\sigma^2}}e^{3 a^2 \Phi/\sigma^2},
 \end{equation}
 where $\mbox{Erf}[\cdot]$ is the error function.
 
 By putting Eq.(\ref{eq1}) into the r.h.s. of the above equation, it becomes obvious that $R_{a}$ is a functional of the LV ratio $b$:
 \begin{equation}
 R_{a}[b\left(\vec{r}\right)] = 1-\mbox{Erf}\left[a \sqrt{\frac{6}{b\left(\vec{r}\right)}}\right]+2 a \sqrt{\frac{6}{\pi b\left(\vec{r}\right)}}e^{-6a^2/b\left(\vec{r}\right)}
 \label{eq2}
 \end{equation}
 Note that  $R_{\mbox{\scriptsize 1}}$ is identified with  the local evaporation rate $R_{\mbox{\scriptsize ev}}$.
 In this case, $R_{a}$ indicates the rate of the particles which spread infinitely.
 When $a < 1$, on the other hand, $R_{a}$ includes  not only the particles escaping infinitely
 but also those trapped in the finite region. 
 For larger value of $a$,
 the particle can move far away from the gravitational center, since $R_{a}$ includes the particles with 
 the higher value of kinetic energy against the absolute value of the
 local potential. Therefore, the value of  $a$
 offers the lower bound of the typical scale $L_{a}$ to which the particle at $\vec{r}$ can move away maximally
 in the future.
 As the local temperature increases against gravitational potential,
 the particle activities are enhanced and 
 the value of $R_{a}$ increase for all of the value of $a$.
 Hence we call $R_{a}$ gravitational fugacity, since it
 represents the particle activities against gravitational potential.
 From the fact that $R_{a}$ is the functional of $b$ in Eq.(\ref{eq2}), LV relation $b\left(\vec{r}\right)=1$ 
 is induced from the global virial condition and the condition that $R_{a}$  is constant everywhere
 (See Appendix \ref{appendix}).
 This means that the SGSs self-organize themselves so as to  make the particle activities constant everywhere.
 In the next section, we will numerically investigate $R_{a}$ for $a=1/2$, since it is the typical value measuring the particle activities against gravitational potential. We will show that the LV relation is attained for the spherical shell where the gravitational fugacity
 exceeds the critical value $R_{1/2}[b=1]$. 
 
 \subsection{ The relation between  the LV ratio $b$  and the gravitational fugacity for anisotropic Gaussian velocity distribution }
 \label{subsec:anisoGauss}
 
 In the above derivation of LV relation, we assume that the velocity 
 distribution is locally isotropic everywhere during the violent relaxation
 process. However, it is well known that velocity dispersion becomes
 anisotropic just after a cold collapse \cite{Albada82}. Hence it seems important
 to examine how strongly the gravitational fugacity  depends on the anisotropy.
 Here in order to ascertain this, we obtain the gravitational fugacity
  for anisotropic Gaussian velocity distribution, where the velocity dispersion
 in tangential direction is different from that in radial direction.
 In the cylindrical coordinates for velocity space, the phase-space
 density can be described as,
 \begin{equation}
 f(v_r,v_t, \vec{r}) = 
 \sqrt{\frac{2}{\pi}} \frac{v_t\rho(\vec{r})}{\sigma_t^2(\vec{r})\sigma_r(\vec{r})}
 \exp\left(-\frac{v_t^2}{\sigma_t^2(\vec{r})}-\frac{v_r^2}{2\sigma_r^2(\vec{r})}\right),
 \label{fanisogaussv}
 \end{equation}
 where $v_r$ and $v_t$ are radial and tangential velocity component
 and $\sigma^2_r$ and $\sigma^2_t := \sigma_\theta^2+\sigma_\phi^2 $ are velocity dispersion of
 radial and tangential component, respectively.
 The gravitational fugacity  for this function form can be evaluated as
 \begin{equation}
 R_{a} \left(\vec{r}\right) = {1 \over \rho} \int\!\!\!\int_{v^2_r+v^2_t\geq a^2 v^2_{\mbox{\tiny cr}}}f(v_r,v_t, \vec{r}) dv_rdv_t.
 \label{aniso-evap}
 \end{equation}
 Substituting (\ref{fanisogaussv}) into (\ref{aniso-evap}) and using the anisotropy parameter defined as
 \be
 \beta := 1-\frac{\sigma_t^2}{2\sigma_r^2},
 \ee
 the fugacity can be described as the functional of both $b$ and $\beta$ as
 \be
 R_{a} \left[b,\beta\right]=1-\frac{2}{(1-\beta)\sqrt{\pi}}\left(\frac{3-2\beta}{3}\right)^{3/2}\int^{a \sqrt{6/b}}_0 x^2 \exp\left(-\frac{3-2\beta}{3(1-\beta)}x^2\right)G\left[\frac{\beta(3-2\beta)}{3(1-\beta)}x^2\right]dx,
 \label{eq2aniso}
 \ee
 where $G\left[ y \right] := \int^1_{-1}\exp\left( y t^2 \right) dt$,
 which correctly reduces to  (\ref{eq2}) when $\beta=0$.
 
\begin{figure}[h]
   \begin{center}
     \begin{tabular}{c c}
 \resizebox{80mm}{!}{\includegraphics{\pathmix/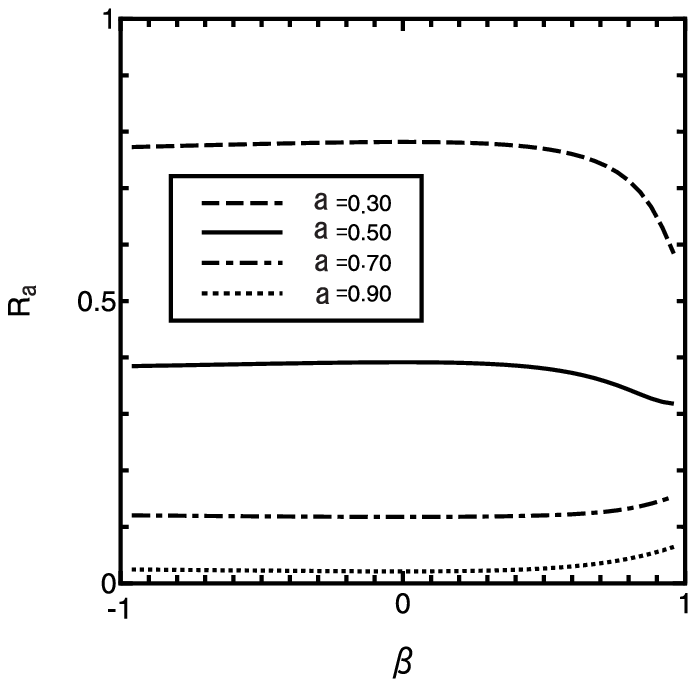}}  &
 \resizebox{80mm}{!}{\includegraphics{\pathmix/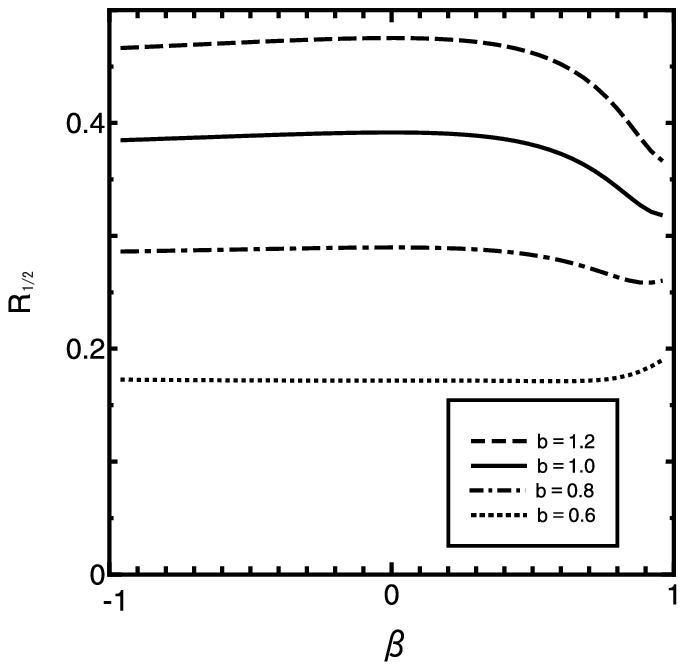}}  \\
 (a) & (b) \\
     \end{tabular}
     \caption{(a) $R_{a}$ for LV ratio $b=1$  as   a function of 
  $\beta$. Each line represents the case with
  $a=0.3$(dash),$0.5$(solid),$0.7$(dot-dash),$0.9$(dot).  For all of the values of $a$, $R_{a}$ is almost
  constant except for the high value of $\beta$.
 (b) $R_{a}$ for  $a=1/2$  as   a function of 
  $\beta$. Each line represents the case with
  $b=0.6$(dot),$0.8$(dot-dash),$1.0$(solid),$1.2$(dash).  For all of the values of $b$, $R_{1/2}$ is almost
  constant except for the high value of $\beta$.
 }\label{evaprate}
   \end{center}
 \end{figure}

 As is shown in Fig.\ref{evaprate}, $R_{a}$ 
 both for the fixed value $b=1$ and for the fixed value of $a$ is almost constant as a function of $\beta$ except for the larger value of $\beta$.
 In our numerical results, the velocity dispersion is not
 highly radially anisotropic  
 at least within a half-mass radius \cite{Sota04,Osamu06}.
 Hence we can roughly say that the gravitational fugacity
 is mainly  determined by  $b$ value in the bound region,
 and the fluctuation
 is minimized in the LV relation even if we take account of
 the anisotropy of velocity dispersion.
 
 \section{Self-organized relaxation process induced from gravitational fugacity for N-body cold collapse.}
 \label{sec:fugacity-demo}
 
 As is shown in our previous papers \cite{Sota04,Sota05,Osamu06}, LV relation is attained 
 for several classes of   cold collapse simulations.
 Here in this section, we will first overview  the results
 of N-body simulations with spherical initial conditions. 
 We will see that the LV ratio oscillates around the critical
 value $b=1$ then converges to it for the collapse of 
 initial homogeneous sphere.
 For the initial cuspy density profile, on the other hand,
 the LV ratio in the central part retains the lower value.
 
 We will also see that such differences among initial conditions can
 be explained by the behavior of gravitational fugacity which
 we defined in the previous section.
 For cold collapse simulations from a homogeneous sphere, we will see that the fugacity
 passes over the critical value   at the moment of maximum collapse. 
 For the collapse of cuspy density profile, on the other hand, the central part of bound region is 
 not so activated that the fugacity  keeps lower than the critical value.
 For N-body simulations we will use the unit of $G=M=r_s=1$, where $M$ and $r_s$ 
 are the total mass and the radius of initial sphere, respectively. The initial free-fall time $t_{ff}:=\sqrt{r_s^3/GM}$ will be used as the time unit for the time-sequence
 of physical variables.

 
 \subsection{Realization of LV relation for several Cold collapse simulations}

\begin{figure}[h]
   \begin{center}
     \begin{tabular}{c c}
 \resizebox{80mm}{!}{\includegraphics{\pathmix/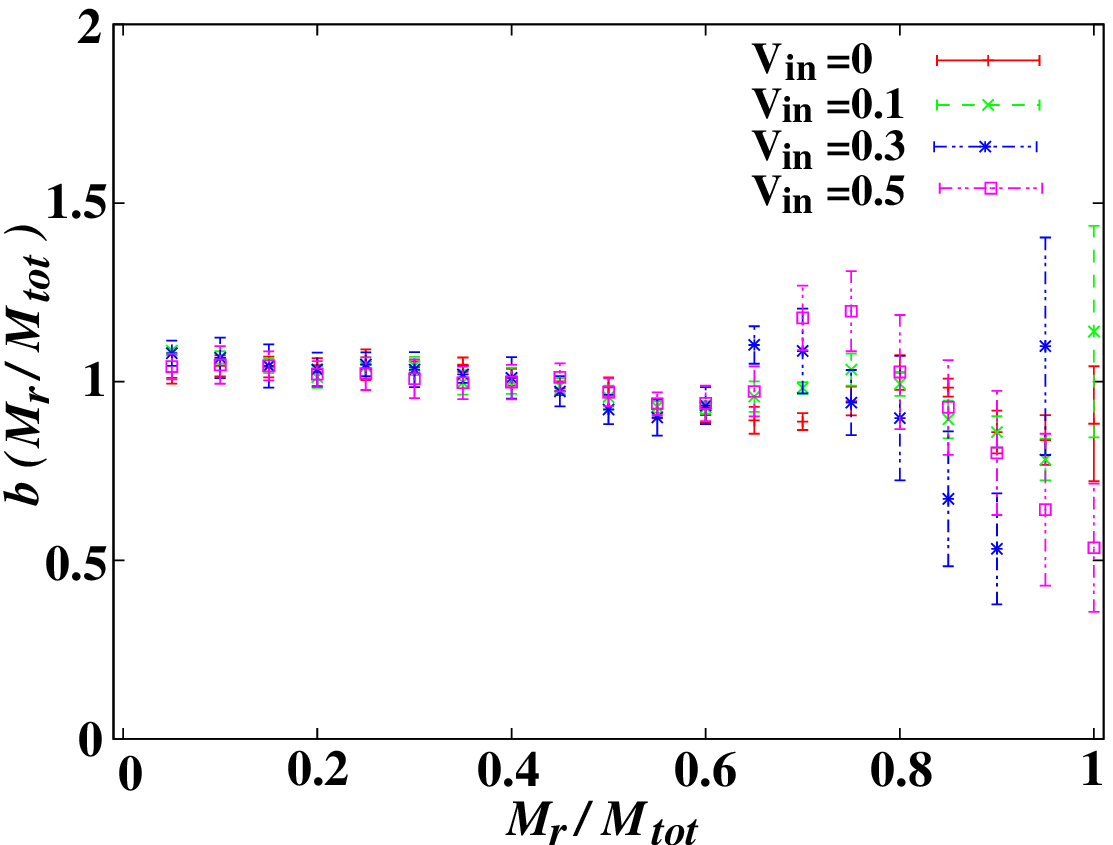}} &
 \resizebox{80mm}{!}{\includegraphics{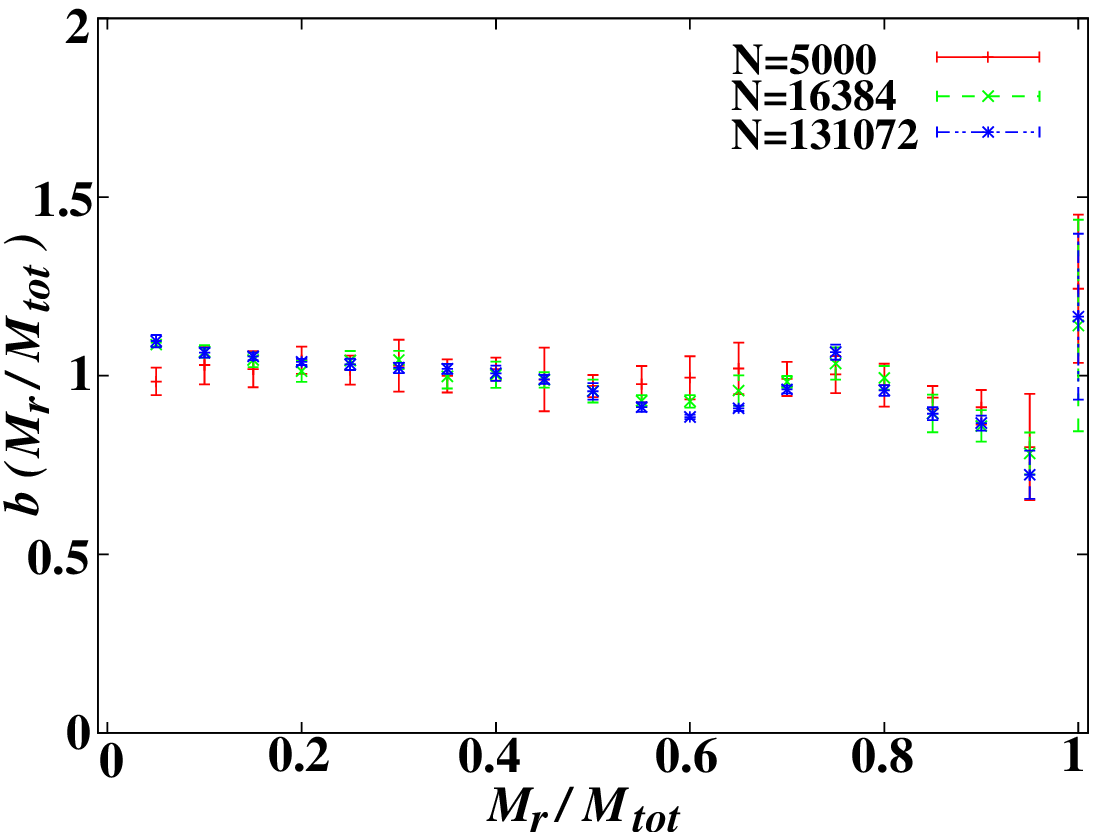}}  \\
 (a) & (b)\\
 \resizebox{80mm}{!}{\includegraphics{\pathmix/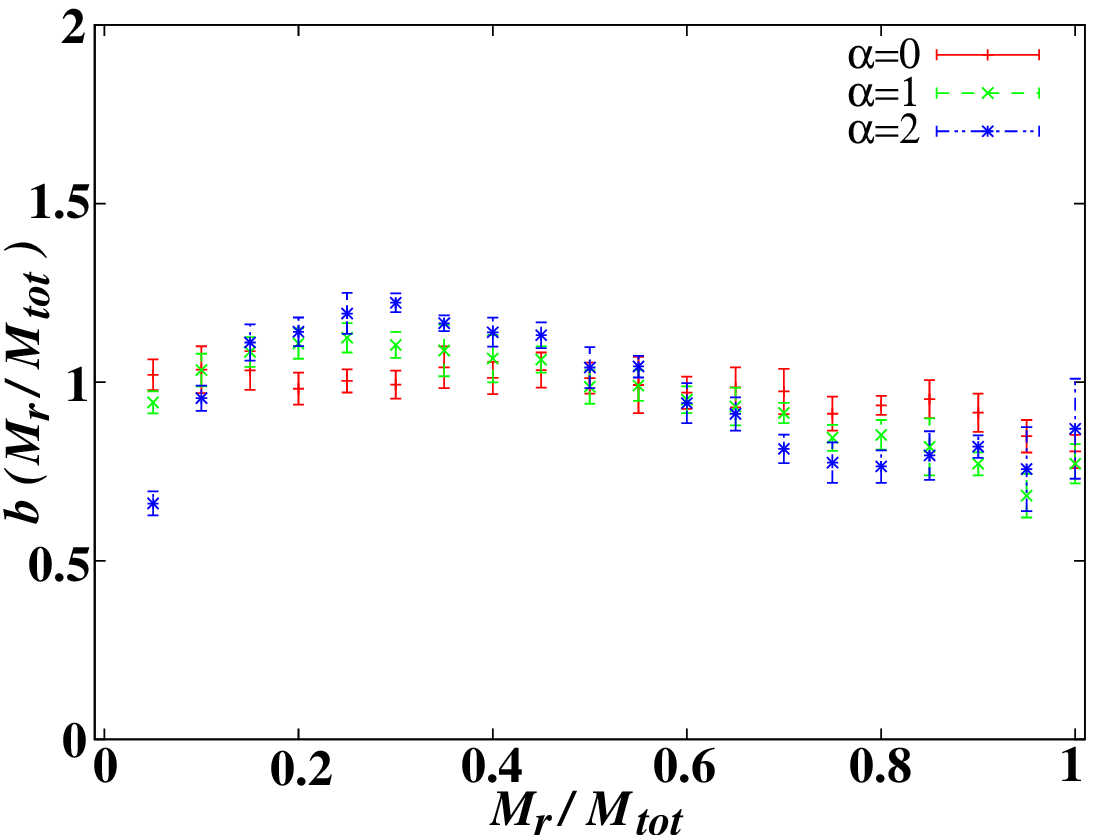}}&
 \resizebox{80mm}{!}{\includegraphics{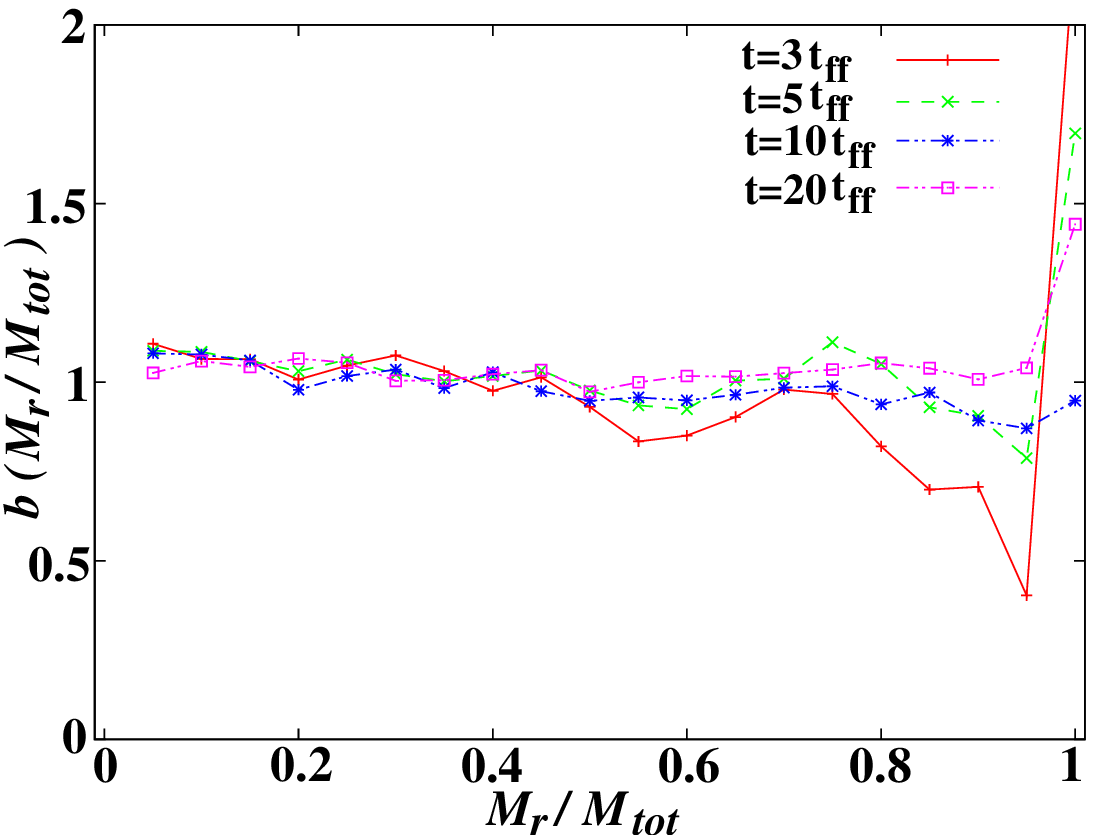}}    \\
 (c) & (d)\\
    \end{tabular}
     \caption{ (a) LV ratio averaged between $t=5t_{ff}$ and $t=10t_{ff}$
     derived from  homogeneous sphere with $N=5000$ and initial global  virial ratio $V_{in}=0.0,0.1,0.3,0.5$. (b) The same as (a) but with $N=5000,2^{14}(=16384), 2^{17}(=131072)$ and $V_{in}=0.0$. (c)The same as (a) but from  spherical
 density profile $\rho \propto r^{-\alpha}$  with exponent
 $\alpha=0$,$1$ and $2$. (d) Snapshots of the LV ratio from  homogeneous sphere with $N=2^{17}$ and $V_{in}=0.0$. Four snapshots
 at $t=3,5,10$ and $20t_{ff}$ are depicted.
  }
     \label{comp-LV-sphere}
   \end{center}
 \end{figure}

 In previous papers \cite{Sota04,Sota05,Osamu06}, we showed that LV relation is well realized for
 the homogeneous cold collapse simulations.
 Here we classify the initial conditions into several categories.
 As well as the previous papers,
  we divide the  bound region composed of the particles 
 with negative energy into plural concentric shells and measure
  the averaged value  of LV ratio on each shell
 between  $t=5t_{ff}$ and  $10t_{ff}$ for the
 collapse from homogeneous sphere (Fig.\ref{comp-LV-sphere}(a),(b),(c)).
  Fluctuations of LV ratio between the period are
  depicted 
 as the error bar in these figures, which 
 represents the r.m.s of LV ratios
 between the time interval. 
 We can see that 
  the LV ratio  converges to the critical value $b=1$ quite well
  in the
 central part of bound region up to $t=10t_{ff}$, while it oscillates 
 around it in the outer part,
 especially for the warmer collapse with
 higher initial virial ratio (Fig.\ref{comp-LV-sphere}(a)).
 
 When we increase the particle number 
 for cold collapse simulations with lower initial virial ratio,
 we can see that LV ratio 
  deviates from the critical one up to  $t=10t_{ff}$,
 although it is not fluctuated (Fig.\ref{comp-LV-sphere}(b)).
 We can speculate that 
 this deviation comes from the  radial
 instabilities for cold collapse characterized as the spiral arms in phase-space \cite{Merrall03}. As the authors showed in the paper,
  such a spiral structure is not stable but rather dissipative.
 In fact, we can numerically ascertain that
 the LV ratio approaches the critical value as the time elapses
 (Fig.\ref{comp-LV-sphere}(d)).
 
\begin{figure}[h]
   \begin{center}
     \begin{tabular}{c c}
 \resizebox{80mm}{!}{\includegraphics{\pathmix/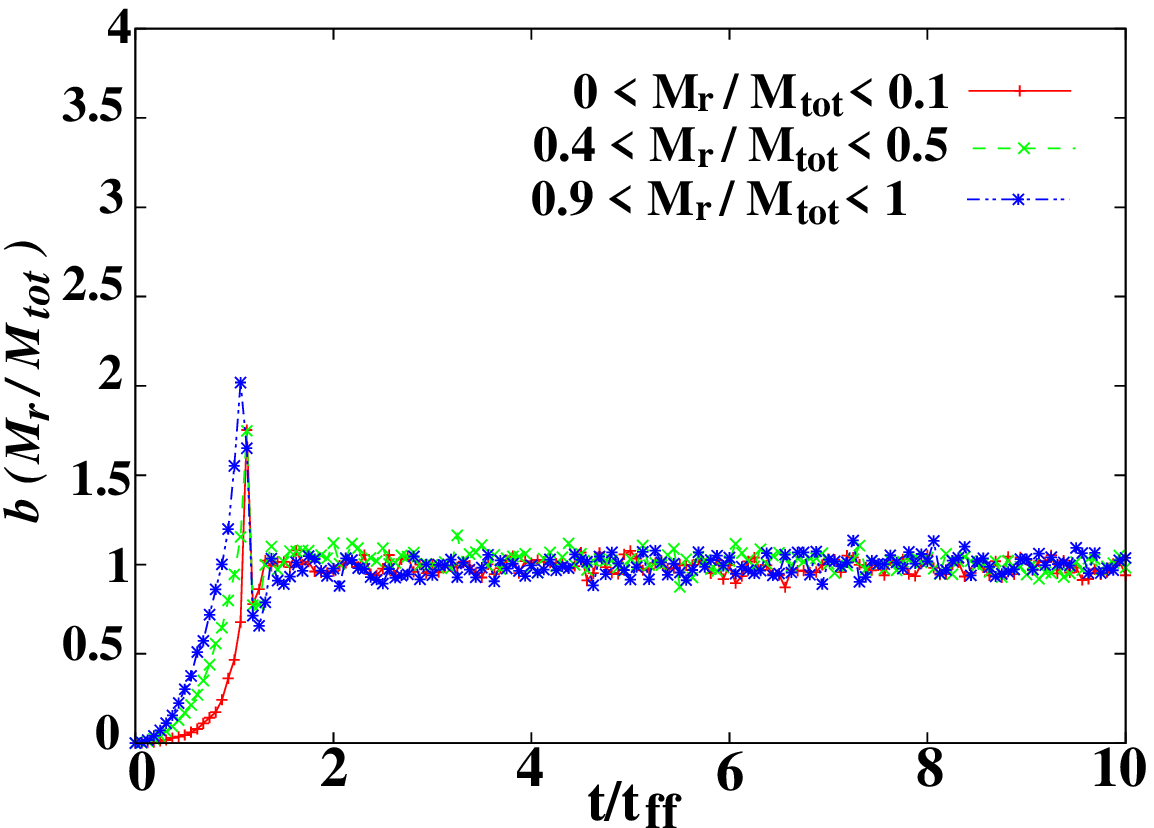}}  &
 \resizebox{80mm}{!}{\includegraphics{\pathmix/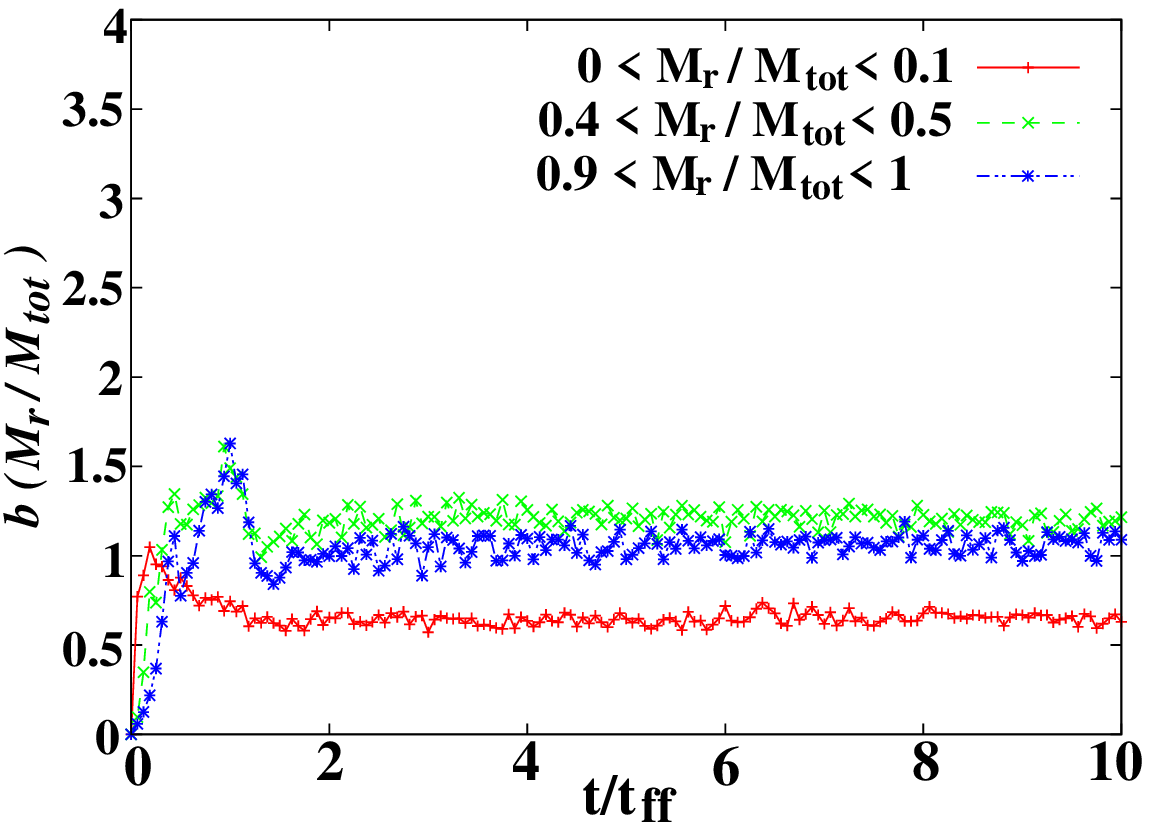}}  \\ 
 (a) & (b)\\  
  \end{tabular}
     \caption{Time sequence of  LV ratio $b$ on fixed shell for cold collapse simulations with $N=5000$ and $V_{in}=0.0$
 (a)for homogeneous  density profile and (b) for cuspy density profile
 with $\alpha=2$. For both of the figures, the bound region is divided into
 $10$ shells and the $b$ value on $1$st, $5$th and $10$th shells are depicted.
  }
     \label{corr-exedcr}
   \end{center}
 \end{figure}

 For the simulations of power-law density profiles,
 on the other hand,
  LV ratio keeps the lower value for a long period
 especially  in the central
 part of the bound region  for steeper density profile (Fig.\ref{comp-LV-sphere}(c)). 
 In fact, 
 we can see that the LV ratio does not pass over the critical value
 for the shells inside 0.1 $M_{tot}$
 even at the moment of the maximum collapse (Fig.\ref{corr-exedcr}(b)). This
 is a remarkable difference from the case with initial homogeneous sphere,
 where  LV ratio  passes over the critical value
 everywhere at the moment of the maximum collapse
 and LV relation is attained (Fig.\ref{corr-exedcr}(a)).

 \subsection{Time evolution of local fugacity for N-body simulations starting from homogeneous sphere}
 
 Here we pay our attention to the time-evolution of  gravitational fugacity for
  several cold collapse simulations because it is directly connected
 with the LV ratio, as is shown in the previous section.
 
\begin{figure}[h]
   \begin{center}
     \begin{tabular}{c c }
 \resizebox{80mm}{!}{\includegraphics{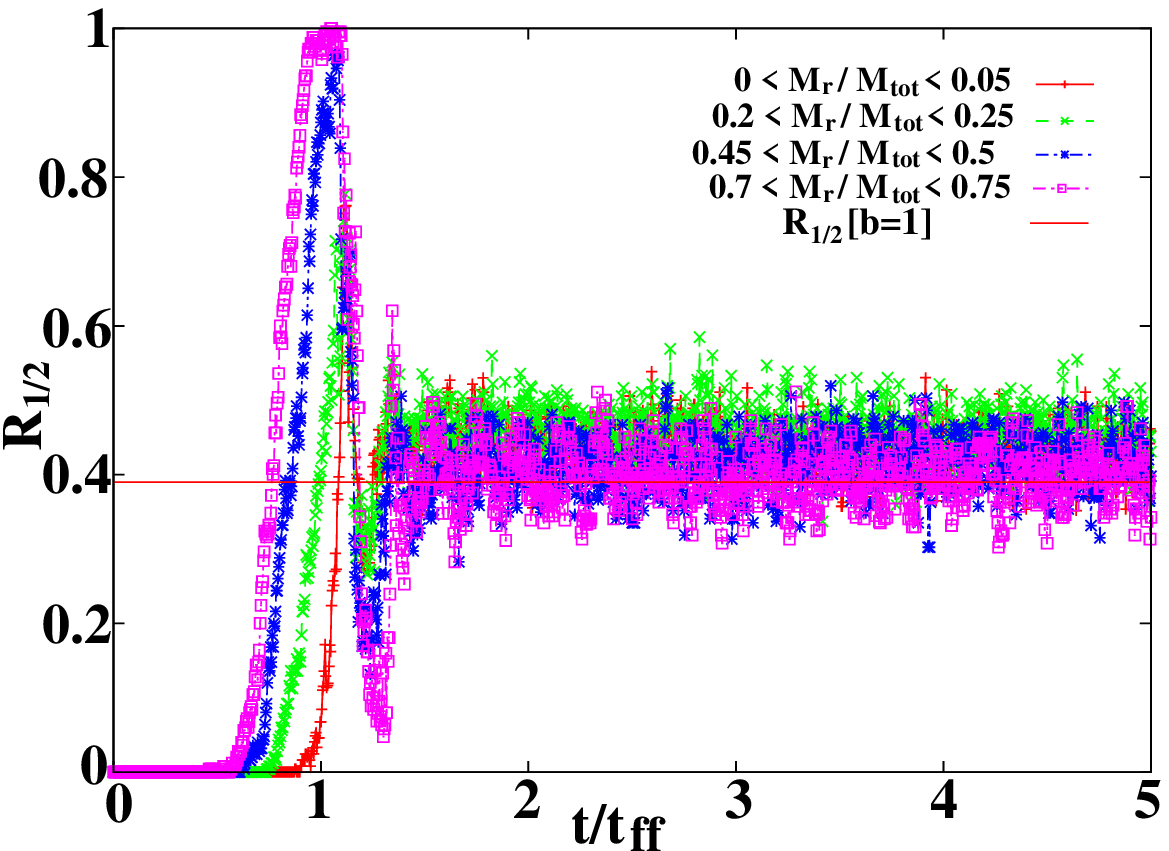}}  &
 \resizebox{80mm}{!}{\includegraphics{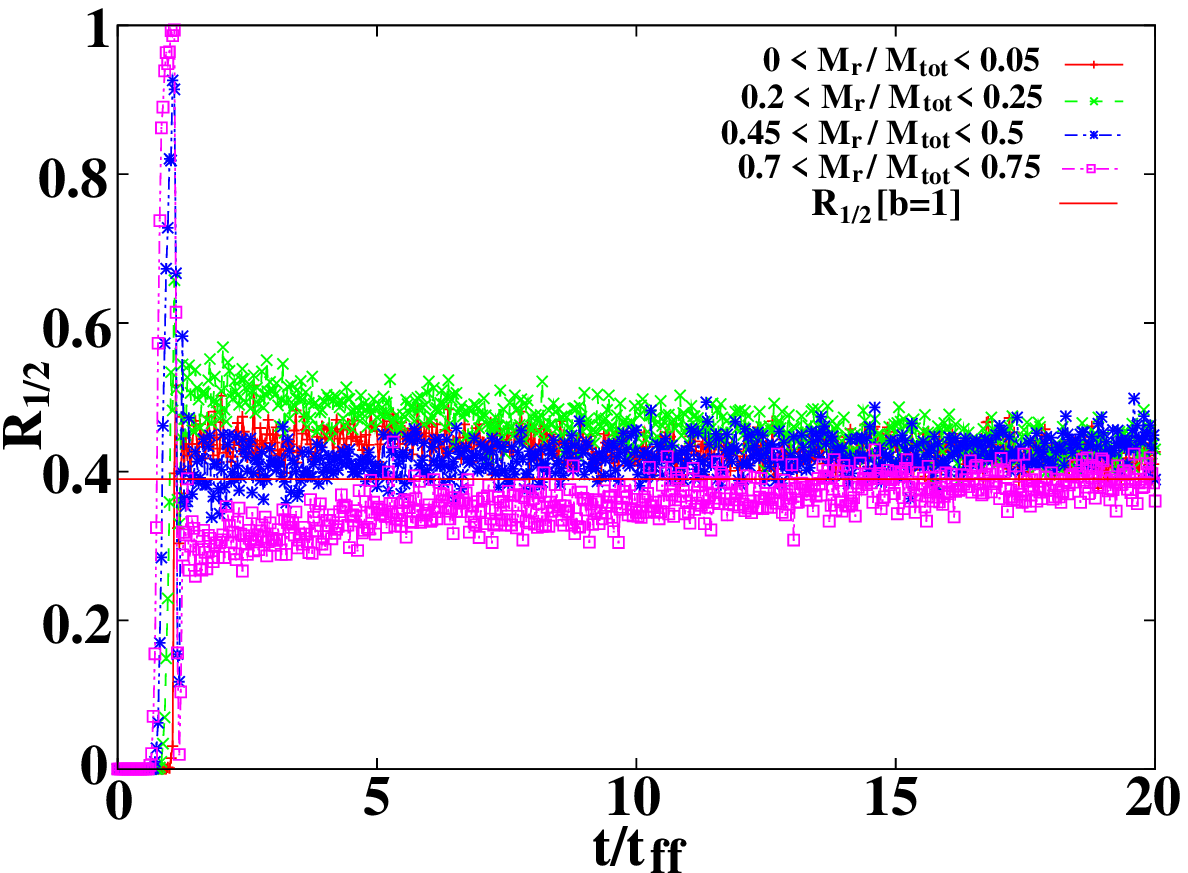}}    \\
 (a) & (b)\\
 \resizebox{80mm}{!}{\includegraphics{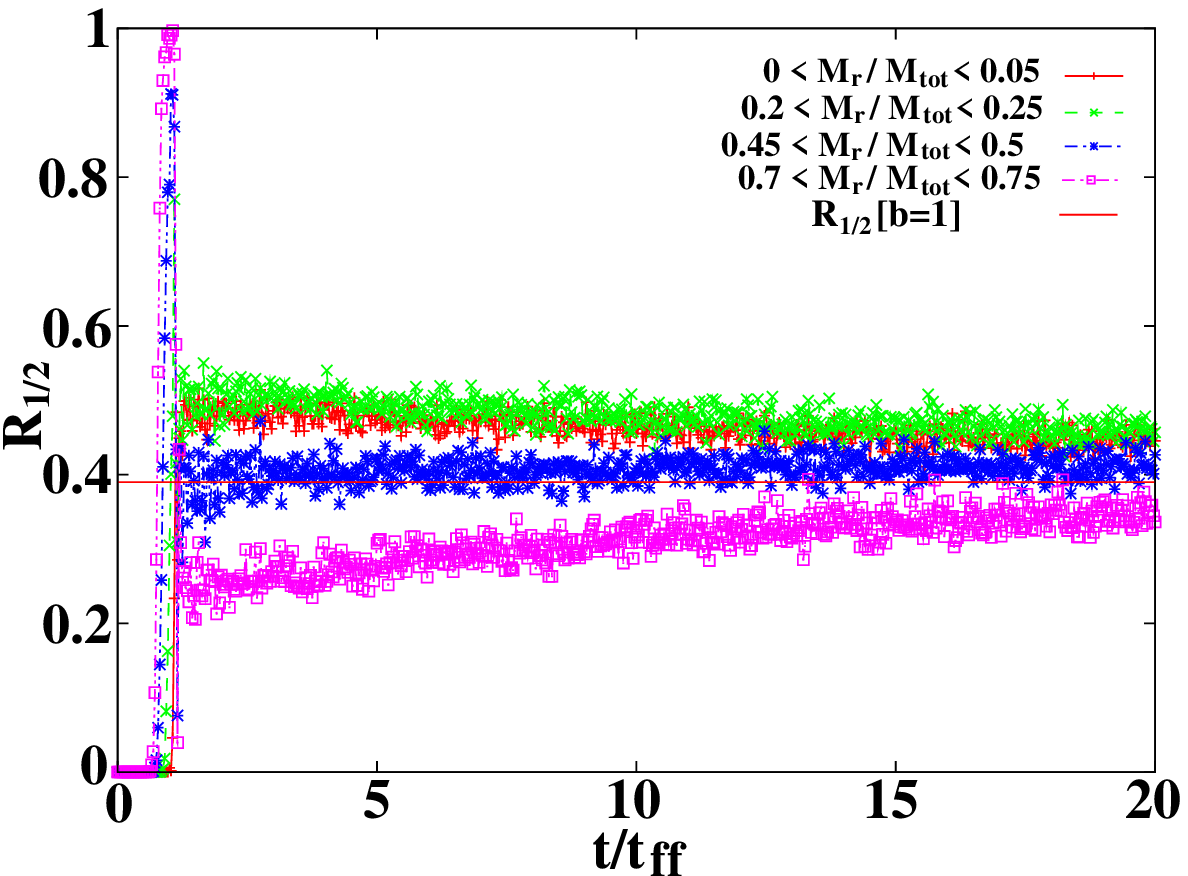}}    &
     \\
 (c) &  \\
 
   \end{tabular}
     \caption{Time evolution of gravitational fugacity $R_{1/2}$
 on each shell for the N-body simulations starting from a homogeneous sphere. Each line represents the value on the shell with the mass ratio
 $M_r/M_{tot}=0.05$(red),$0.25$(green),$0.5$(blue),$0.75$(pink) with (a)$N=5000$ ,(b) $N=2^{14}(=16384)$,(c) $N=2^{15}(=32768)$. The horizontal red line represents
 the critical value $R_{1/2}[b=1]$.}
     \label{comp-fug-homo-Npar}
   \end{center}
 \end{figure}

 We 
 can estimate the value of fugacity $R_{1/2}$ simply by counting the number
 of particles whose velocity exceeds the value $0.5 v_{cr}$ on
 each shell. For the case starting from a homogeneous sphere
 with vanishing virial ratio $V_{in}
 =0$,
 $R_{1/2}$ passes over  the critical value $R_{1/2}[b=1]$
 at the moment of maximum collapse
 on all of the shells (Fig.\ref{comp-fug-homo-Npar}). Once it passes over, it turn out to be reduced,
 because the particles are too activated to stay there and turn to 
 spread out against gravitational potential.
 This seems plausible, because
 just after the moment of maximum collapse,
 the particles efficiently begin to spread out from inner region to outer
 by climbing up the potential, because of their high kinetic energies.
 This state with excessive kinetic energy  can  be  quantified  as
 the high value of gravitational fugacity. 
 Since they lose the kinetic energy as they go up the potential hills, the local fugacity turn out to be reduced.
 If the averaged fugacity becomes lower than the averaged value,
 they begin to increase. Hence the oscillations of fugacity occur
 until the local fugacity at each position are balanced with each other.
 Finally it settles down to the distribution consistent with the LV relation $b=1$.
 
 We can see that the convergence to the critical value 
 is very fast for all of the shells in the case with smaller particle number (Fig.\ref{comp-fug-homo-Npar}(a)), while
 it becomes slower
 for the outer shell in the case with larger particle number
  (Fig.\ref{comp-fug-homo-Npar}(c)), which is because of the 
  surviving of a shock created at the maximum collapse.
  Therefore such a deviation fades down as the time elapses.
  
\begin{figure}[h]
   \begin{center}
     \begin{tabular}{c c }
 \resizebox{80mm}{!}{\includegraphics{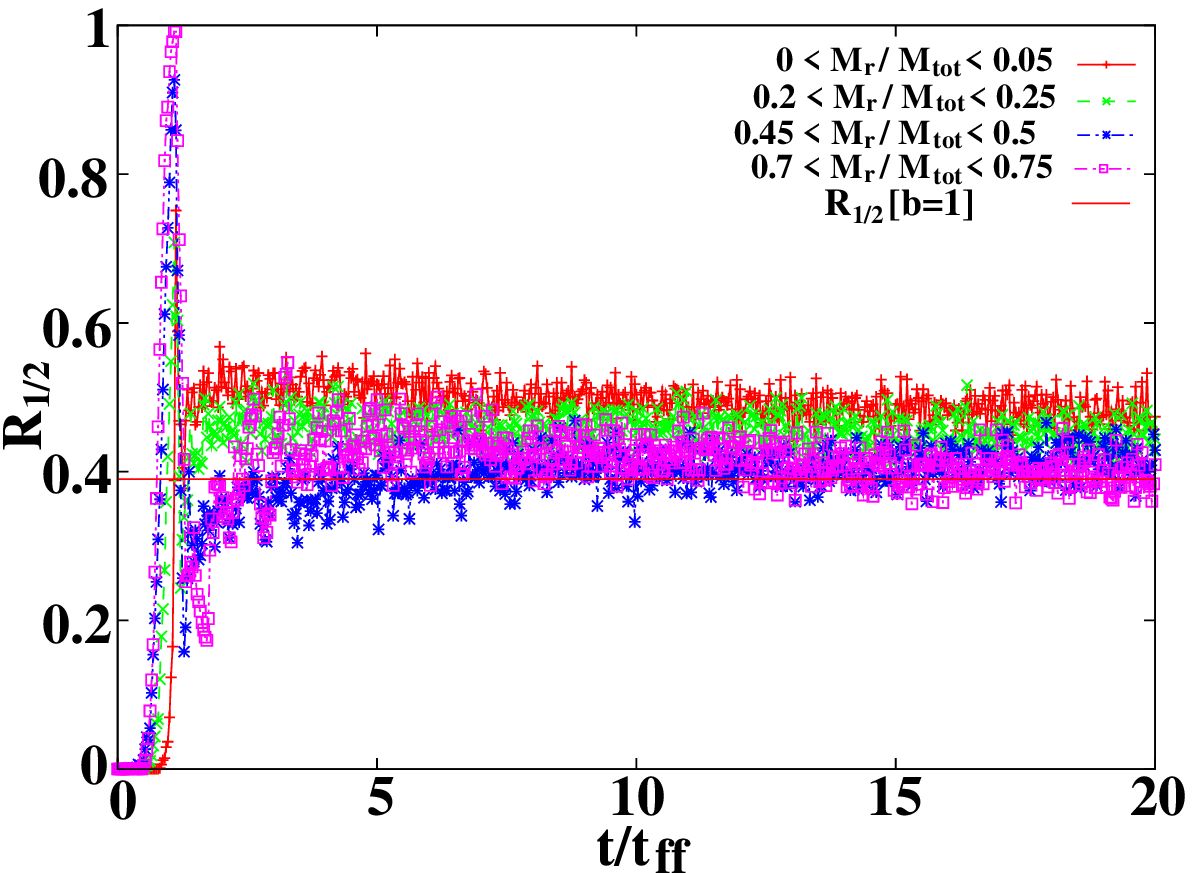}}  &
 \resizebox{80mm}{!}{\includegraphics{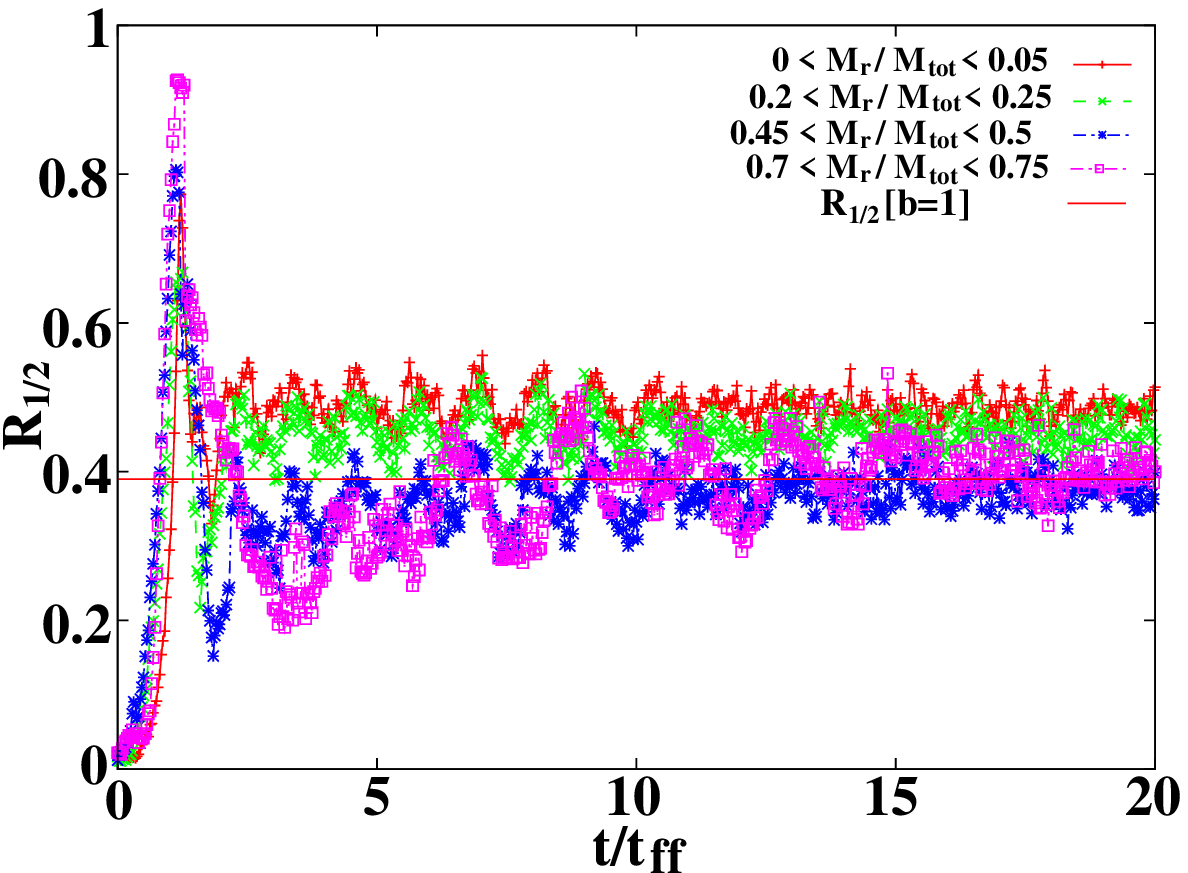}}    \\
 (a) & (b)\\
 \resizebox{80mm}{!}{\includegraphics{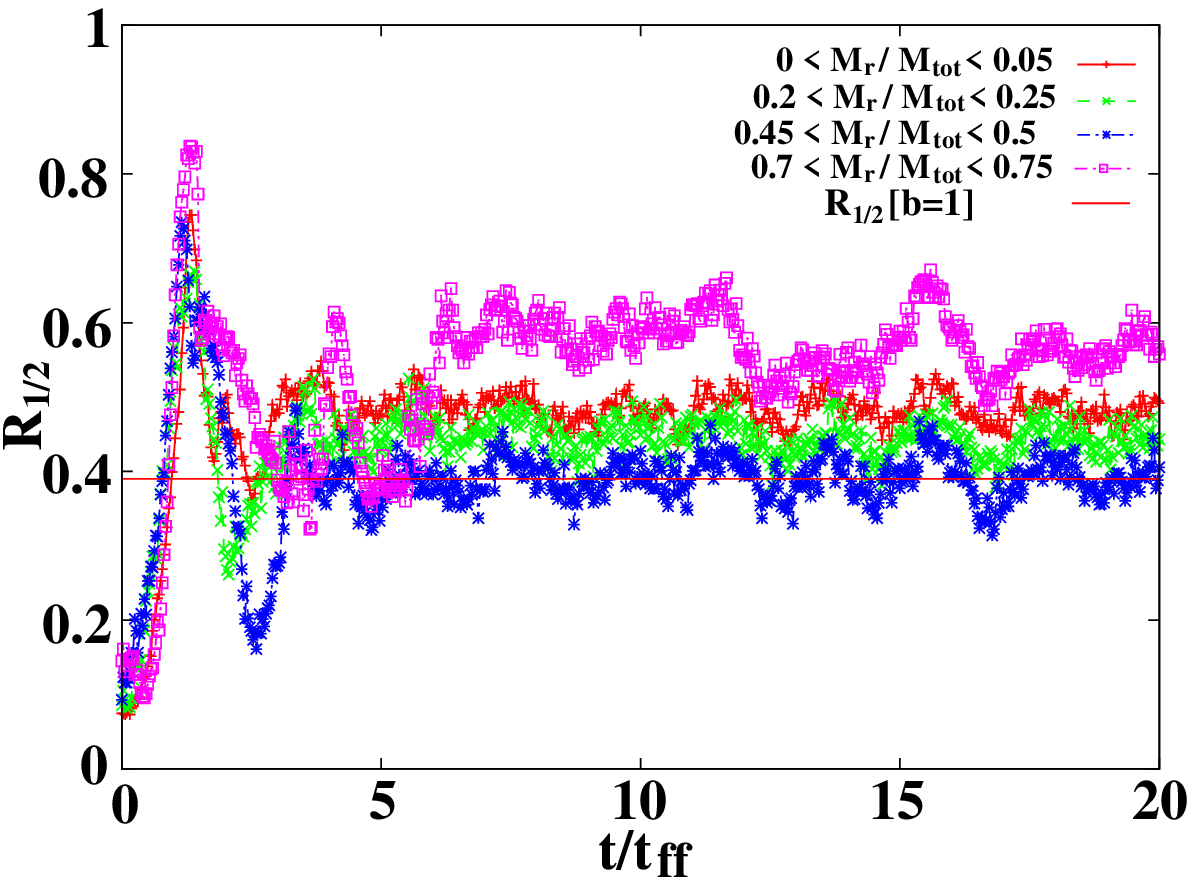}}  &
      \\
 (c) &  \\
   \end{tabular}
     \caption{Same as Fig.\ref{comp-fug-homo-Npar} but for the fixed
 particle number $N=2^{14}$ and for different initial virial ratios
 with $V_{in}=\mbox{(a)}0.1,\mbox{(b)}0.3,\mbox{(c)}0.5$. }
     \label{comp-fug-homo-virial}
   \end{center}
 \end{figure}

 We can also investigate  the warmer initial conditions with initial virial ratio $V_{in}=0.1,0.3,0.5$
 (Fig.\ref{comp-fug-homo-virial}).
 As the initial virial ratio is higher, fugacity on each shell
 oscillates with a longer time-interval. This is because 
 the density of central core becomes lower for warmer collapse.
 Since the  time scale of the oscillation is determined by the free-fall
 time determined by the value of central density,
 we need to wait longer for a warmer collapse until it
 settles down to the stationary state.

 \subsection{Time evolution of local fugacity for N-body simulations  starting from a cuspy density profile}
 
 Next we will investigate the initial cuspy density profile,
 where LV ratio becomes less than one in the innermost part of the shell,
 even after the system settles down to the stationary state.
 
\begin{figure}[h]
   \begin{center}
     \begin{tabular}{c c }
 \resizebox{80mm}{!}{\includegraphics{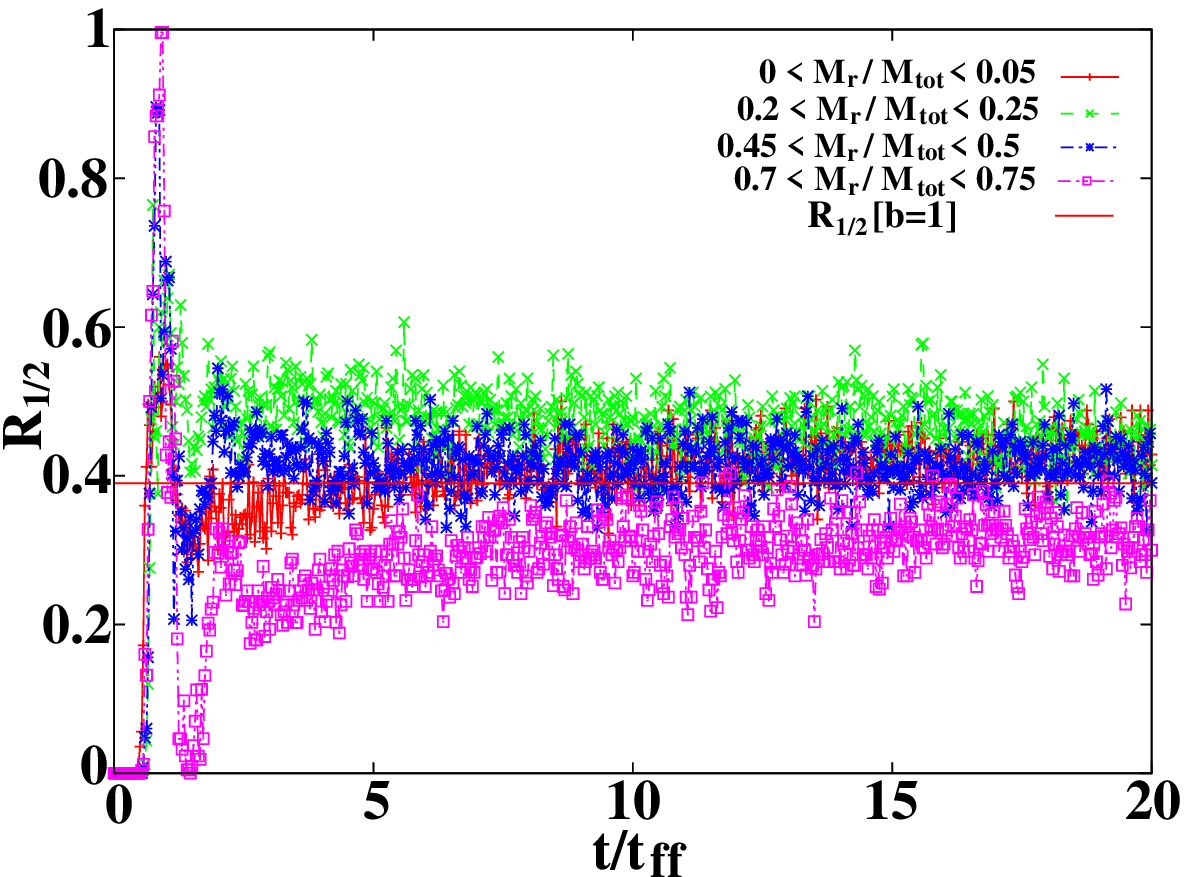}}  &
 \resizebox{80mm}{!}{\includegraphics{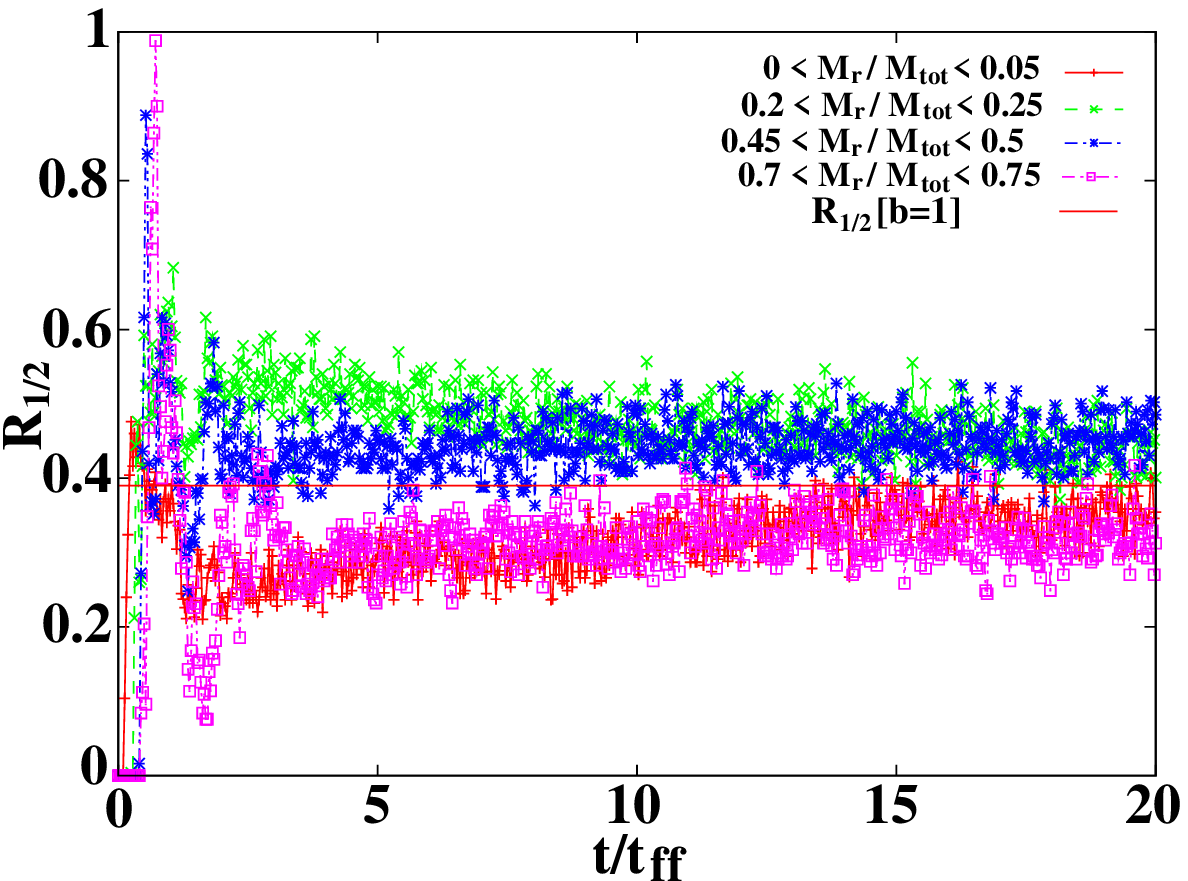}}    \\
 (a) & (b)\\
 
   \end{tabular}
     \caption{Same as Fig.\ref{comp-fug-homo-Npar} but for the initial
     density profile $\rho \propto r^{-\alpha}$  with exponent
 $\alpha=$ (a)$1$ and (b) $2$. }
     \label{comp-cusp-fugacity}
   \end{center}
 \end{figure}

 Here we will compare the time-evolution of fugacity on several shells.
 Gravitational fugacity certainly synchronizes with LV ratio $b$ and 
 keeps lower in the central part (Fig.\ref{comp-cusp-fugacity}).
 The fugacity at the innermost shell does not pass over the critical
 value $R_{1/2}[b=1]$, neither does LV ratio (Fig.\ref{corr-exedcr}).
 We also checked that the $R_{a}$ does not pass over the critical
 value $R_{a}[b=1]$ for $a<0.5$.
 This means that the particles in the central
 region do not spread out to the outer region but stay around the center,
 which prevents  any part of the bound region  from realizing the LV relation.
 This state with excessive potential energy in the central part
 is highly stable, since it is located in the central core 
 and isolated from the outer part.

 \section{The character of self-organized equillibrium state from the viewpoint of LV condition}
 \label{sec:fugacity-demo2}
 
 In Sec.\ref{sec:fugacity-def}, we found that local gravitational fugacity  depends on the local position only through 
 the LV ratio $b$. Hence we can naturally judge
 the activity of the local region through its LV ratio $b$. 
 That is, if some parts of the region take $b$ lower  than $1$,
 their activities are less than the averaged value, while if
 they take $b$ higher than $1$, they are more activated.
 
 For the initial cuspy density profile with
 steeper configuration the central part of the system is not driven to 
 the state with LV relation, but is
 trapped to the quasi-equilibrium state, 
 because it stays in the state with $b<1$.
 This state may
 or may not be stable under the perturbation in
 phase-space. 
 In cosmological simulations, this cuspy density profile has  common
 characters with the equilibrium state attained
  as well as the 
 lower temperature and power-law phase-space density \cite{Taylor01,Dehnen05}.
 In such cosmological settings, there a large number of such cuspy halos
 moving and interacting each other. Since each of them is influenced
 by the gravitational forces from other nearby halos,
 it is worth examining the robustness
 of such a quasi-equilibrium state which does
 not satisfy the LV relation.
 Hence, in this section, we examine the stability of both
 the LV equilibrium state and the 
 quasi-equilibrium state especially from the
 viewpoints of the LV relation.

 \subsection{Self-adjusting mechanism through the enhancement of activities for the state with $b<1$}
 
\begin{figure}[h]
   \begin{center}
     \begin{tabular}{c c}
 \resizebox{80mm}{!}{\includegraphics{\pathmix/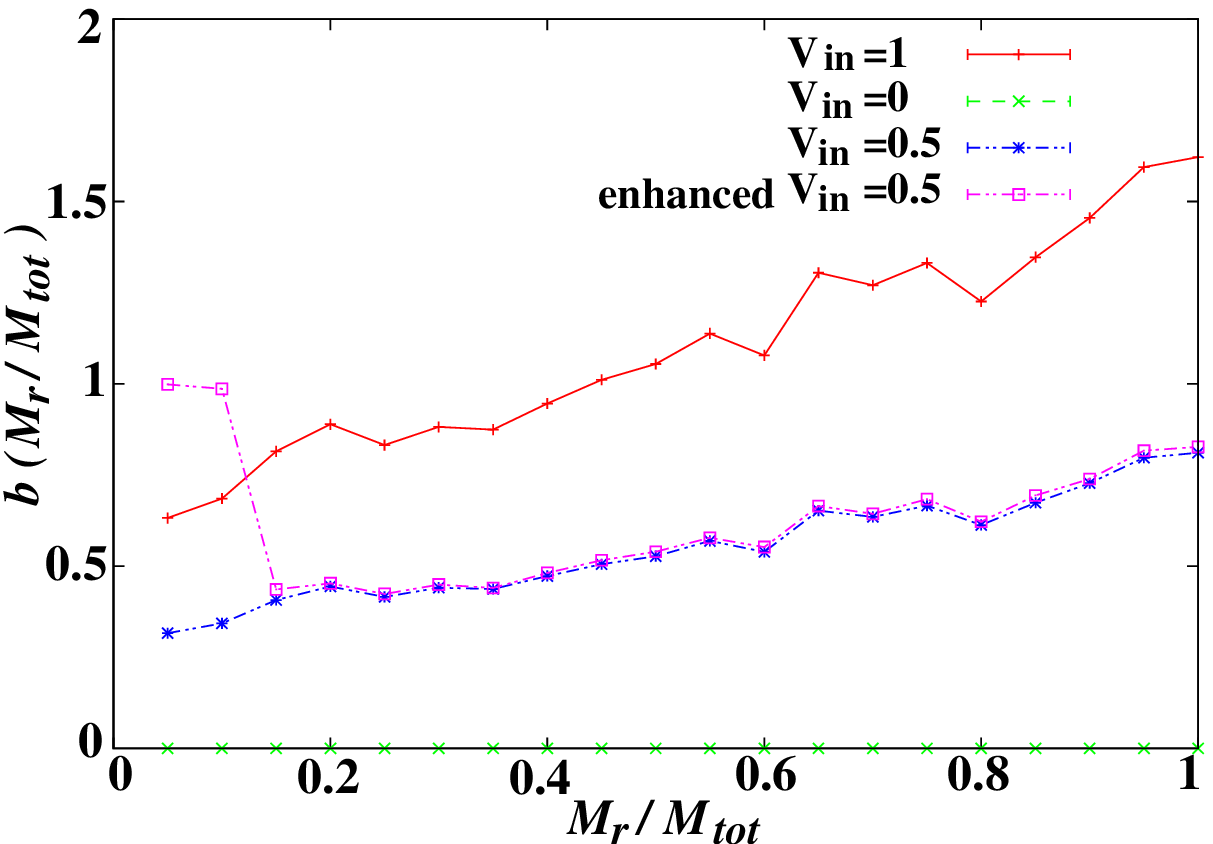}}    &
 \resizebox{80mm}{!}{\includegraphics{\pathmix/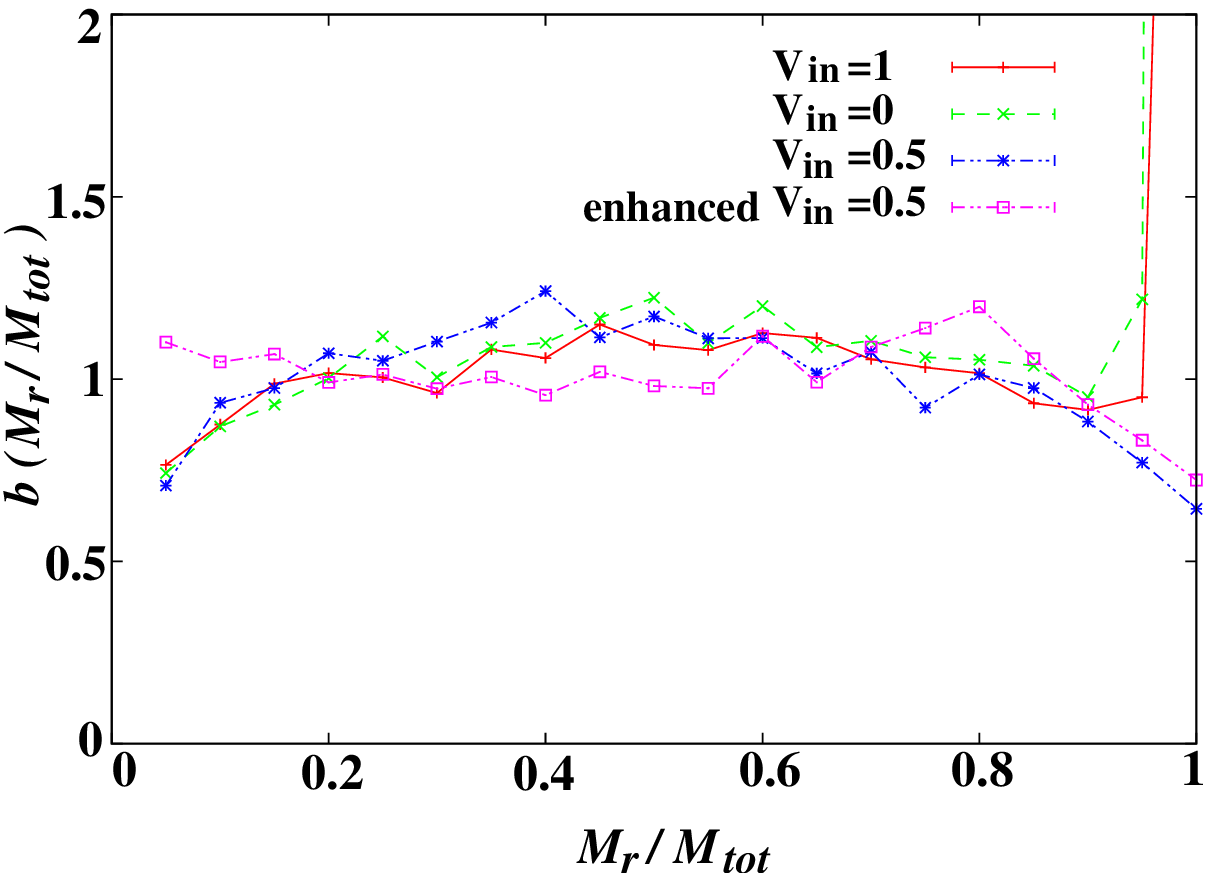}} \\ (a)  &    (b)\\   \end{tabular}
     \caption{Snapshots of the $b$ distribution as a function
 of $M_r$ obtained from the cold collapse simulations
 with $N=5000$ and $V_{in}=0.0$
   for initial   cuspy density profile
 with $\alpha=2$
  at  $t=0$(a)  and $80t_{ff}$ (b). The pink line represents the case that the initial $b$ value in the inner most shell is enhanced to $b=1$. }
     \label{evolve-delta-cusp}
   \end{center}
 \end{figure}

 First, we examine a cuspy  density profile with 
 the exponent  $\alpha=2.0$ for several 
 initial virial ratios ($V_{in}=0.0,0.5,1.0$).
 As is shown in Fig.\ref{evolve-delta-cusp}, $b(M_r)$ takes the minimum value
 at the center of mass and monotonically increases toward
 outside for  these initial distributions.
 Hence the particle activities becomes lower and lower toward
 the center of mass.
 Starting from these initial conditions Fig.\ref{evolve-delta-cusp}(a)), we can get the
 equilibrium state where the central part of $b(M_r)$
 is lower than the critical value (Fig.\ref{evolve-delta-cusp}(b)).
 During this process, the outer part particles
 tend to spread out because of the high activities while the inner
 parts tend to collapse because of the lower activities.
 Hence they do not mix themselves but keep their inner part isolated
 from the outer part.

 On the other hand, if we enhance the LV ratio for the inner most shell up to one, we can enhance the activity of particles against gravitational
 potential.
 In this case, not only at the central shell but also at several shells
 in the inner part, LV ratio 
 exceeds  the critical value  at some future moment,
 which lead to LV critical state almost everywhere in 
 the bound region  (Fig.\ref{evolve-delta-cusp}(b)).  
 
 From these empirical results, the cuspy density profile
 turns out to be unstable against the fluctuations which
 activate the central part of the bound region.
 From such activations, the system begins to evolve
 toward the critical state satisfying LV relation.
 This means that the state of the cuspy density profile 
 is pre-saturated and can be evolved to the saturated 
  state satisfying LV relation, against the
 perturbation which causes strong particle activities in the central
 part.

 \clearpage
 
 \subsection{Robustness of the LV critical state against LV
 perturbations}
 
\begin{figure}[h]
   \begin{center}
     \begin{tabular}{c c}
 \resizebox{80mm}{!}{\includegraphics{\pathmix/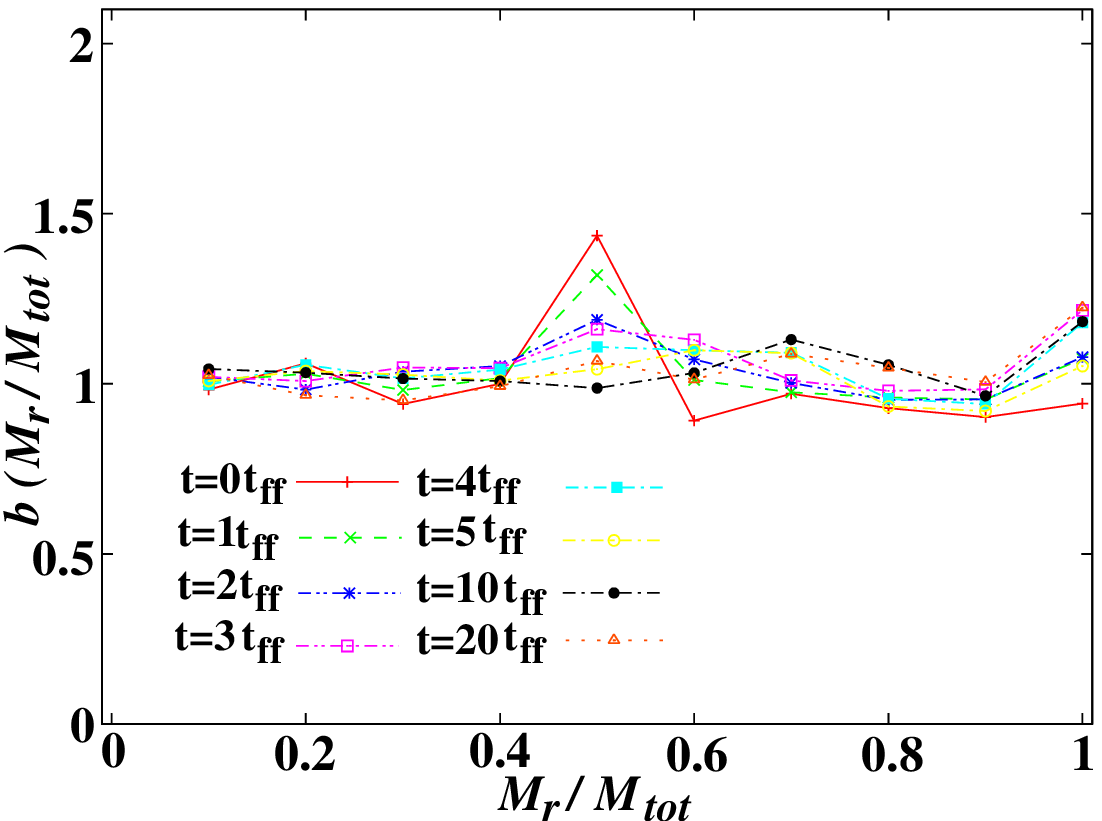}} &
 \resizebox{80mm}{!}{\includegraphics{\pathmix/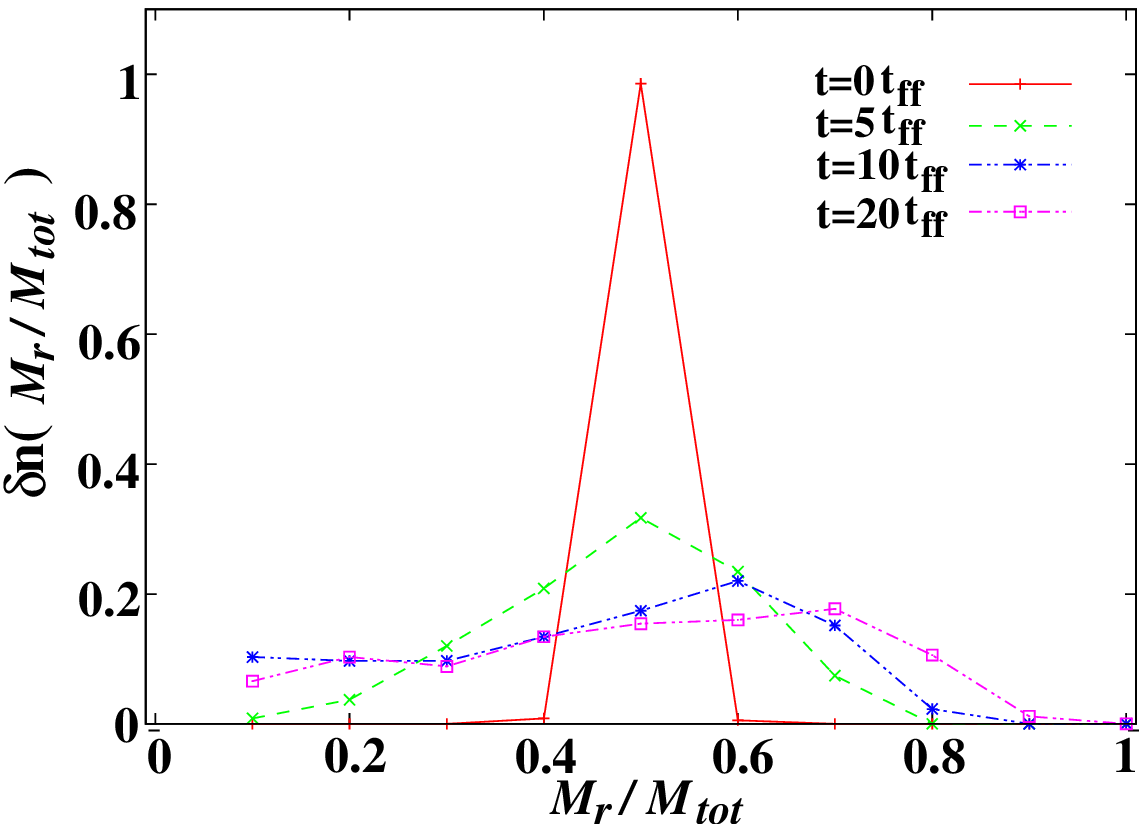}} \\
 (a) & (b) \\
 \resizebox{80mm}{!}{\includegraphics{\pathmix/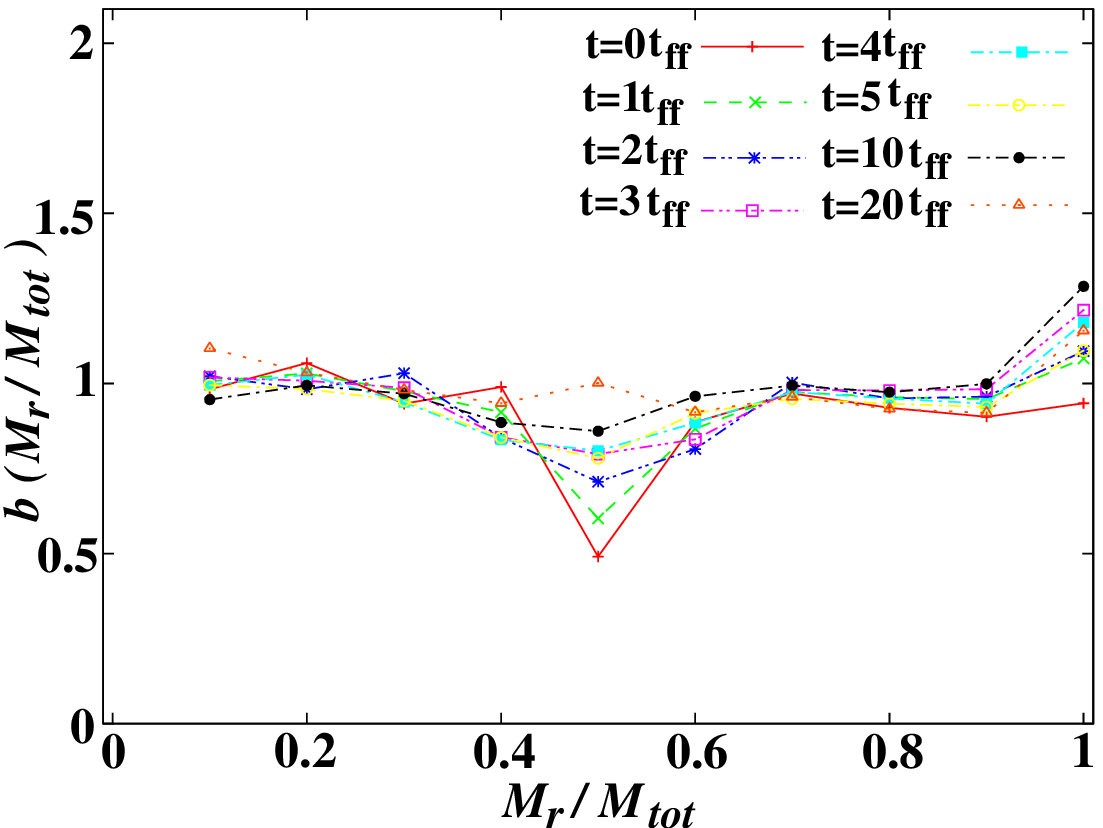}} &
 \resizebox{80mm}{!}{\includegraphics{\pathmix/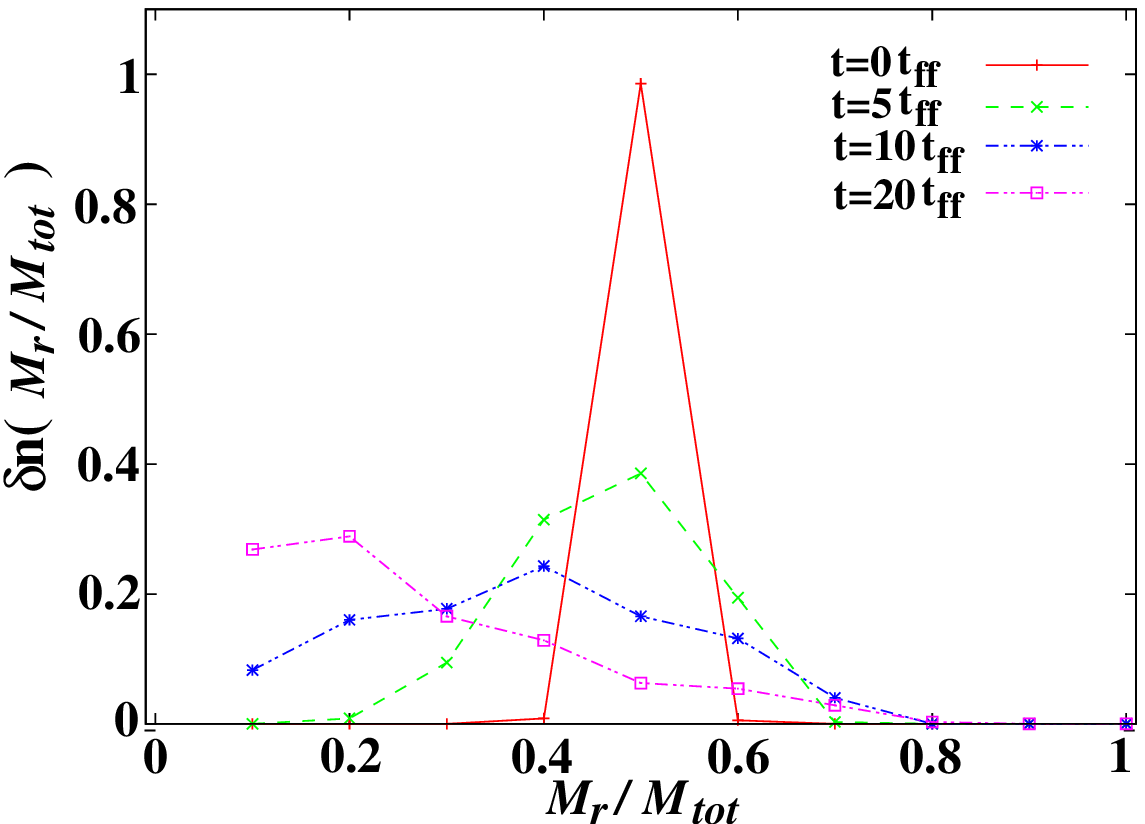}} \\
 (c) & (d) \\
     \end{tabular}
     \caption{The time evolution of LV distribution after the LV ratio of one shell  is artificially
 shifted from the critical value after the system reaches the LV state. 
 Here the LV ratio around $M_r=0.5M_{tot}$ is shifted   from  upper (a) and lower (c) for the equilibrium state 
  starting from a homogeneous sphere with  $V_{in}=0.0$ and with $N=5000$.
 The system smoothly goes back to the critical state $b=1$ within $10t_{ff}$ time interval.
 The distribution of perturbed particles $\delta n$ at several moments 
 are  also depicted in (b)  and in (d) for the upper and lower
 shift of $b$, respectively.
 The particles are smoothly spread toward other shells for both of the cases.
 }
     \label{equi-shift}
   \end{center}
 \end{figure}

 Next,  we examine the stability of the LV equilibrium
 state against the deviations from $b=1$.  Here in order to see this character,
  we shift the $b$ value of one selected shell for the LV 
 critical state
 by increasing or decreasing the velocities for the particles in the critical shell 
 at the same rate. This means that the activity of each shell is
 slightly different from the averaged one.
 Then we can examine the dissipation
 of the particles by  tracing their positions in later time (Fig.\ref{equi-shift}).
 
 When we increase the LV ratio $b$, the fugacity $R_{a}$ is enhanced
 and the particles are expanded to the outer shell and the system settles down to the state with $b=1$ everywhere (Fig.\ref{equi-shift}(a)).
 On the other hand, when we decrease the LV ratio, the fugacity is reduced
 and the particles fall down to the inner shell and the system again goes
 back to the state with $b=1$ everywhere (Fig.\ref{equi-shift}(b)).
 Hence the critical state with $b=1$ is stable against the shift of the
 local activities.

 \section{Concluding remarks}
 \label{sec:conclusions} 
 
 In this paper, we propose the gravitational fugacity  which
 represents the activity of particles against gravitational potential.
 We also showed that the gravitational fugacity  at each local region is determined directly
 by the LV ratio at the region.  Hence we can pose the physical meaning
 on the LV relation as the condition that the local activity of particles
 are balanced with each other on any local region.
 In fact, we showed that LV ratio $b$ oscillates around the critical value $b=1$  and settles down to it when LV relation is achieved.
 Especially we found that the LV relation is attained when LV ratio
 passes over the critical value $b=1$ everywhere
 in the bound region at least once before
 the system reaches stationary equilibrium state.
 Hence the SGS is self-organized toward  the critical
 state with $b(\vec{r})=1$ under the condition that
 the system passes over the critical state $b=1$ everywhere.
 The gravitational fugacity $R_{a}$ synchronizes with $b$ and
 converges to the critical value $R_{1/2}[b=1]$, when LV ratio
 passes over the critical value $b=1$.

 In cold collapse process starting from a  homogeneous sphere, the SGS
 starts from a state with  low  fugacity everywhere. Both LV ratio and   fugacity pass over the critical value largely everywhere at the moment of the initial  collapse, which causes violent relaxation.
 On the other hand, in  cold collapse process starting from a cuspy density profile, LV ratio does not pass over the critical value in
 the central part of the bound region and 
 most particles stay there until the system reach the quasi-equilibrium state. In this case, the gravitational fugacity $R_{a}$ with
 $a \leq 0.5$ does not pass over the critical value 
 $R_{a}[b=1]$.
 This is because most particles in the central part falls down
 into the deep  potential  well induced from the outside particles.
 In this case the system stays in the low fugacity state and never
 reaches the state with LV relation.
 Hence we can divide the collisionless relaxation process into two
 categories; one admitting that the LV ratio passes over the
 critical value $b=1$ everywhere
 in the bound region and another which does not.
 The former self-organizes itself 
 so as to minimize both the LV ratio and the local fluctuation of  the gravitational fugacity  as much as possible.
 
 This self-organized process with the condition $b>1$  reminds us of the self-organized criticality (SOC)
 in sandpiles \cite{self-org}.
 In sandpiles, avalanches occurs whenever the slope exceeds the critical value.
 In the self-organized process in SGS, particles are well activated before it
 reaches the state with LV relation.
 Such activated particles tend to spread out effectively, which may correspond
 to the avalanche in sandpile.
 Hence it seems worth examining that  the character of SGS self-organized
 process from the viewpoint of SOC, which is characterized by the scaling
 behavior.
 Such similarities of self-organized process
 with SOC will be well analyzed in our upcoming paper.

 In our previous paper  \cite{Osamu04}, we showed that the quasi-equilibrium state
 attained through cold collapse simulations in SGS can be characterized
 as the velocity distribution superposed with Gaussian distribution
 with different local velocity dispersion corresponding to the
 local temperature.
 LV relation indicates that the velocity dispersion normalized 
 with the local potential becomes constant and independent of the position.
 Hence, from the viewpoints of statistical mechanics, we can say that
 the system settles down not to the isothermal state
 with constant temperature but to the quasi-equilibrium state
 with constant pseudo-temperature normalized with potential.
 In this critical state, particles can move around the region in
 different temperature and reaches the equilibrium state. Superstatisics
 is proposed as such a model superposing the different temperature
 although they seem to lack it's  theoretical derivation from dynamical process \cite{Beck05}. 
 SGS self-organized relaxation may give the hint for explaining the 
 superstatistics from more fundamental dynamical viewpoints.

 In cosmological simulations, the stable stationary 
 solution exists
 for the Jeans equations under the assumption that
 the $\rho/\sigma^3$ follows the scaling law.
 This solution is special in that it has the property
 that particles spread infinitely but the total mass
 of the bound region is finite \cite{Taylor,Dehnen05}
 In fact, this
 stationary solution can describe the NFW density profile derived from 
 cosmological simulations quite well.
 On the other hand, for the cold collapse simulations,
 the bound state follows the stationary solution
 of the Jeans equation
 under the assumption of LV relation, which also
 has the property of infinitely spread and finite mass \cite{Osamu06,Evans05}.
 So far the global stability of two solutions have  not been
 compared because they are the solutions under
 different constraints.
 Our analysis of gravitational fugacity may
  give a hint for
 the stability of these two sorts of solutions.
 As long as we examined, the cuspy density profile
 which does not follow LV relation is a pre-saturated
 state which goes to the LV state if the central
 isolated region is activated. 
 In actual astrophysical systems, such a transition
 from the  pre-saturated state to the LV stable
 state may happen through
 the sequence of merging process. In fact, we found that the
 system approaches the LV state for merging process, although
 the time scale of relaxation is much longer than those of cold collapse
 simulations.
 The effect of such sequence of merging on the LV criticality remains
 as a future work.

 From astronomical points of view, observing LV relation may give the hint for merging history of dark matter halos around galaxies.
 In fact, An \& Evans proposed the general form of phase-space distribution function following LV relation and
 utilized them to evaluate the relation between the cusp slope of dark matter halo in Galaxy and the anisotropy of $\gamma$-ray flux radiated from Galactic Center \cite{An05}. Such observations may also reveal how well  the LV relation 
 is attained in dark matter halos around galaxies.

 \section*{Acknowledgements}
 
 Numerical computations were in part carried out on GRAPE system at the Center for Computational Astrophysics, CfCA, of the National Astronomical Observatory of Japan
 (g07b12).

 \begin{appendix}
 \appendix
 
 \section{Global virial relation}
 \label{appendix}
  
  In Eq.(\ref{eq2}), $R_{a}$ is a monotonic
  function of $b$ for any fixed value of $a$. Hence if $R_{a}$ is constant and independent of the position, $b$ also becomes constant. This means that the local velocity dispersion $\sigma^2$  becomes proportional to the local potential everywhere.
  Assuming the spheical symmetry and describing the Eq.(\ref{eq1}) 
  with the cumulative mass $M_r$ as the coordinate in radial direction,
  we obtain the relation
  \begin{equation}
  \sigma^2(M_r)= -b\Phi(M_r)/2.
  \label{app-eq1}
  \end{equation}
  
  Integrating both sides of the Eq.(\ref{app-eq1}) with
  $M_r$  in the full of the bound region, we obtain the relationship between total kinetic energy $K$ and total potential $W$ as
  \begin{equation}
  2K=-b W,
  \end{equation}
  where $K$ and $W$ are the total kinetic energy and total potential
  defined as 
  \bea
  K &=& \frac{1}{2}\int_0^{M_{tot} } {\sigma^2\left( M_r \right)} dM_r \non \\
  W &=& \frac{1}{2}\int_0^{M_{tot} } {\Phi \left( M_r \right)} dM_r,
  \eea
  respectively.
  Hence the LV ratio $b$ is identified with  the global virial ratio $V$ defined as 
  \be
  V=-{2K \over W},
  \label{gvirial}
  \ee
  which means that $b$ becomes equal to one for a
  globally virialized state.
 
 \end{appendix}


 

\begin{thebibliography}{}
 
 \bibitem{Albada82}
 T.~S.~van Albada, Mon.~Not.~R.~Astron.~Soc. \textbf{201}, 939 (1982).
 
 \bibitem{Aguilar91}
 L.~A.~Aguilar and D.~Merritt, ApJ. \textbf{354}, 33 (1990).
 
 \bibitem{Navarro96}
 J.~F.~Navarro, C.~S.~Frenk, and S.~D.~M.~White, ApJ. \textbf{462}, 563 (1996).
 
 \bibitem{Navarro97}
 J.~F.~Navarro, C.~S.~Frenk, and S.~D.~M.~White, ApJ. \textbf{490}, 493 (1997).
 
 \bibitem{Taylor01}
 J.~E.~Taylor and J.~F.~Navarro, ApJ. \textbf{563}, 483 (2001).
 
 \bibitem{Lynden67}
 D.~Lynden-Bell, Mon.~Not.~R.~Astron.~Soc. \textbf{136}, 101 (1967).
 
 \bibitem{Horthy91}
 J~.Horthy and J.~Madsen, Mon.~Not.~R.~Astron.~Soc. \textbf{253}, 703 (1991).
 
 \bibitem{Horthy93}
 J~.Horthy and J.~Madsen, Mon.~Not.~R.~Astron.~Soc. \textbf{265}, 237 (1993).
 
 
 \bibitem{Sota04}
 Y.~Sota, O.~Iguchi, M.~Morikawa, and A.~Nakamichi, astro-ph/0403411(unpublished).
 
 \bibitem{Sota05}
 Y.~Sota, O.~Iguchi, M.~Morikawa, and A.~Nakamichi,
 Prog.~Theor.~Phys.~Suppl. \textbf{162}, 62 (2006).
 
 \bibitem{Osamu06}
 O.~Iguchi, Y.~Sota, A.~Nakamichi, and M.~Morikawa, 
 Phys.~Rev.~E \textbf{73}, 046112 (2006).
 
 \bibitem{Binney87}
 J.~Binney and S.~Tremaine, \textit{Galactic Dynamics}
 (Princeton University Press, Princeton, 1987).
 
 \bibitem{Eddington16}
 A.~S.~Eddington, Mon.~Not.~R.~Astron.~Soc. \textbf{76}, 572 (1916).
 
 \bibitem{Evans05}
 N.~W.~Evans and J.~An, Mon.~Not.~R.~Astron.~Soc. \textbf{360}, 492 (2005).
 
 
 
 \bibitem{Merrall03}
 T.~E.~C.~Merrall and R.~N.~Henriksen, ApJ. \textbf{595}, 43 (2003).
 
 \bibitem{Osamu04}
 O.~Iguchi, Y.~Sota, T.~Tatekawa, A.~Nakamichi, and M.~Morikawa, 
 Phys.~Rev.~E \textbf{71}, 016102 (2005).
 
 
 
 \bibitem{Dehnen05}
 W.~Dehnen and D.~E.~McLaughlin, Mon.~Not.~R.~Astron.~Soc. \textbf{363}, 1057 (2005).
 
 \bibitem{self-org}
 H.~J.~Jensen, \textit{Self-Organized Criticality} (Cambridge University Press 1998).
 
 \bibitem{Beck05}
 C.~Beck and E.~D.~G.~Gohen, Physica A,  \textbf{322}, 267 (2003).
 
 \bibitem{An05}
 J.~An and N.~W.~Evans, Astron.~Astrophys. \textbf{444}, 45A (2005).
 
 \end{thebibliography}
 
 \end{document}